# Overlapping Multi-hop Clustering for Wireless Sensor Networks

WINC-TR-1001

Moustafa Youssef, *Senior Member, IEEE,* Adel Youssef, *Member, IEEE,* and Mohamed Younis, *Senior Member, IEEE*

***Abstract*—Clustering is a standard approach for achieving efficient and scalable performance in wireless sensor networks. Traditionally, clustering algorithms aim at generating a number of disjoint clusters that satisfy some criteria. In this paper, we formulate a novel clustering problem that aims at generating overlapping multi-hop clusters. Overlapping clusters are useful in many sensor network applications, including inter-cluster routing, node localization, and time synchronization protocols. We also propose a randomized, distributed multi-hop clustering algorithm (*KOCA*) for solving the *overlapping* clustering problem. *KOCA* aims at generating connected overlapping clusters that cover the entire sensor network with a specific average overlapping degree. Through analysis and simulation experiments we show how to select the different values of the parameters to achieve the clustering process objectives. Moreover, the results show that *KOCA* produces approximately equal-sized clusters, which allows distributing the load evenly over different clusters. In addition, *KOCA* is scalable; the clustering formation terminates in a constant time regardless of the network size.**

## I. Introduction

In recent years, sensor networks have attracted much interest in the wireless research community as a fundamentally new tool for a wide range of monitoring and data-gathering applications. Sensor nodes are significantly constrained in the amount of available resources such as energy, storage, and computational capacity. Due to energy constraints, a sensor can communicate directly only with other sensors that are within a small distance. To enable communication between sensors not within each other's communication range, sensors form a multi-hop communication network.

Clustering is a standard approach for achieving efficient and scalable performance in sensor networks. Clustering facilitates the distribution of control over the network and, hence, enables locality of communication. Moreover, clustering nodes into groups saves energy and reduces network contention as nodes communicate their data over shorter distances to their respective cluster-heads.

Traditionally, clustering algorithms, e.g. [1]–[27], aim at generating a number of disjoint clusters that satisfy some criteria, e.g. minimum number of clusters. In this paper we formulate a *novel* clustering problem that aims at generating overlapping multi-hop clusters. In overlapping clusters, a node may belong to more than one cluster, in contrast with the traditional clustering algorithms, in which each node belongs to only one cluster. Overlapping clusters are useful in many sensor network applications, including inter-cluster routing [28], node localization [29]–[32], and time synchronization protocols [33]. Fig. 1 shows an example of overlapping clusters in a sensor network. Note that although a node belongs to more than one cluster, this does not imply that the node has to be on all the time. This decision is the responsibility of the application that runs on top of the generated clusters.

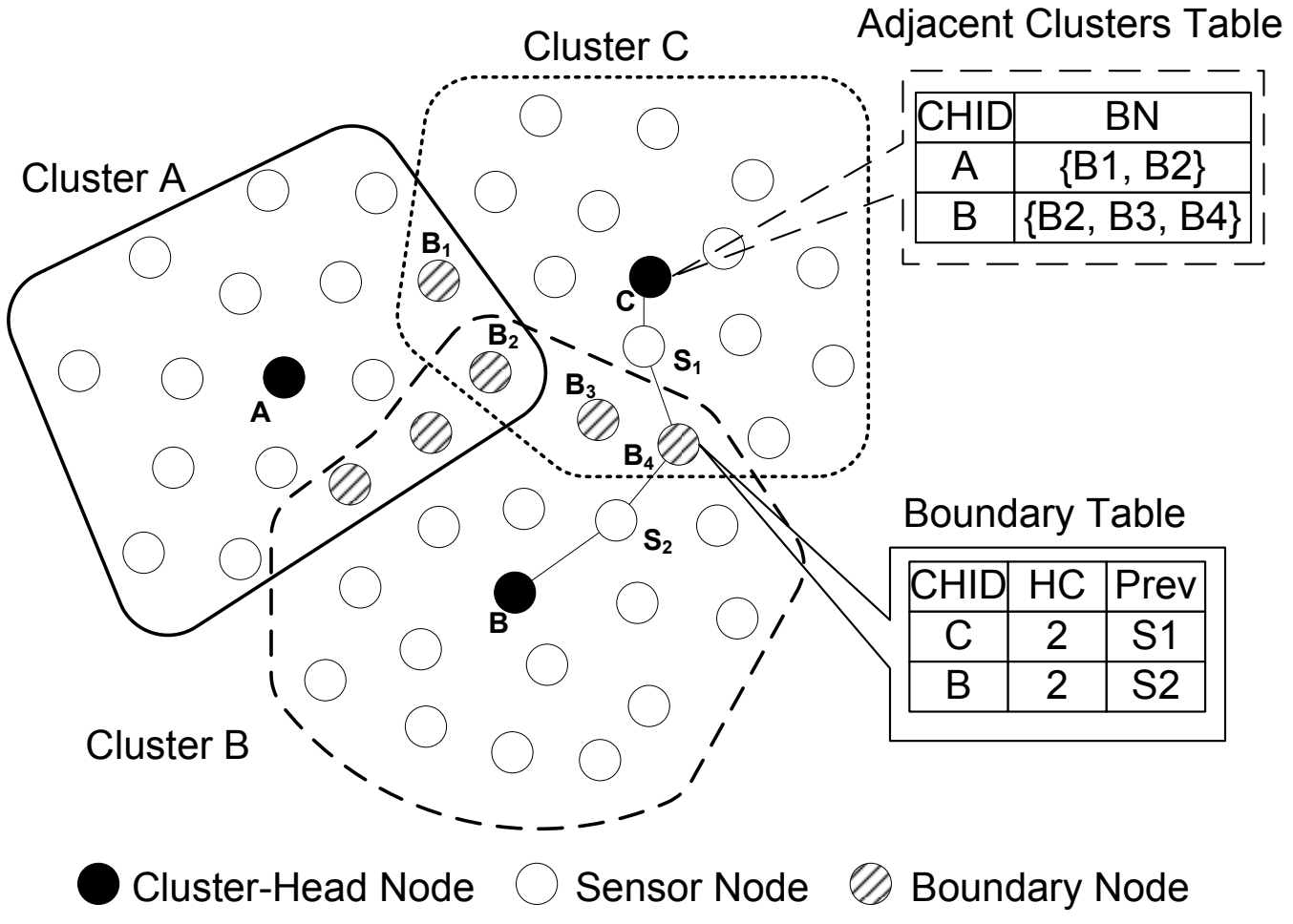

Fig. 1. An example of overlapping clusters. The figure shows the *Boundary Table* at boundary node $B_4$ and the *Adjacent Clusters Table* at cluster-head $C$.

The overlapping multi-hop clustering is a challenging problem to solve in a distributed manner, which is the case in a wireless sensor network. We show that the overlapping multi-hop clustering problem is NP-hard and propose the **k**-hop **O**verlapping **C**lustering **A**lgorithm (*KOCA*) as a randomized distributed algorithm for solving it. *KOCA* aims at generating connected overlapping clusters that cover the entire sensor network with a desired average number of boundary nodes in the overlapping area. *KOCA* operates in stationary networks where nodes do not move and have equal significance, which is a reasonable assumption for sensor networks. Nodes randomly elect themselves as cluster heads with some probability $p$. The cluster head probability ($p$) is a given parameter to the algorithm that can be tuned to control the number of overlapping clusters in the network. After the termination of

M. Youssef is with Nile University, Egypt.
A. Youssef is with Google Inc., USA.
M. Younis is with University of Maryland, Baltimore County, USA.

the clustering process, each node is either a cluster head or within $k$-hop from *at least one* cluster head, where $k$ (*cluster radius*) is another given parameter to the algorithm. The clustering process terminates in $O(1)$ iterations, independent of the network size, and does not depend on the network topology or size.

Through analysis and simulation experiments we show how to select the different values of the parameters to achieve the clustering process objectives. Moreover, the results show that the *KOCA* algorithm incurs low overhead in terms of messages exchanged and produces approximately equal-sized clusters, which allows distributing the load evenly over the different cluster heads. In addition, *KOCA* leads to a low normalized stdev (stdev divided by the mean) of overlapping degree, implying consistent overlapping degree between different clusters.

In summary, the contributions of this work is threefold: (1) we formulate the overlapping k-hop clustering problem for wireless sensor networks, (2) we present a randomized distributed heuristic algorithm for solving the problem, and (3) we evaluate the performance of the proposed algorithm through analysis and simulation.

In the balance of this section we describe some applications for the overlapping k-hop clustering problem and present the paper organization.

### A. Applications of the Overlapping k-hop Clustering Problem

Having overlapping clusters with some degree is beneficial in numerous applications. For example, overlapping clusters can boost the resilience of cluster-based routing protocols, such as [28], [34]–[39], to node failure or compromise. Establishing overlapping clusters leads to providing multiple paths between each pair of overlapping clusters, increasing the network robustness against boundary nodes failures and distributing packet-forwarding load between clusters.

Another application for overlapping clusters is in *anchor-free* localization, e.g. [29]–[32], [40], [41]. Anchor-free localization algorithms try to compute nodes' positions without the use of anchor nodes (i.e. nodes that know their positions). In this case, the algorithm estimates relative positions, in which the coordinate system is established by a reference group of nodes. The network is divided into small clusters of nodes, where each cluster has its own *relative* coordinate system. In order to construct the *global* network topology, we need to map between the different relative coordinate systems. In order to transform between two clusters in 2D, there must be at least three *common boundary nodes* (i.e. the two clusters are overlapping with degree at least three). This transformation is normally performed by the cluster heads. The proposed *KOCA* algorithm can be used in this case to achieve this overlapping degree. Similar concepts are needed for time synchronization where local synchronization within clusters can be mapped into global synchronization through the shared boundary nodes [42].

### B. Paper Organization

The paper is organized as follows. Section II formulates the overlapping k-hop clustering problem. In Section III, we present the *KOCA* heuristic algorithm for solving the problem. We analyze the performance of the *KOCA* algorithm in Section IV. Section V provides extensive simulation experiments for evaluating the *KOCA* algorithm and validating the analysis. In Section VI we survey related work. Finally, Section VII gives concluding remarks.

## II. Overlapping Multi-hop Clustering

### A. System Model

We consider a multi-hop homogeneous wireless sensor network where all nodes are alike[1]. We assume that each node has a unique id. In addition, the nodes are location-unaware, i.e. not equipped with GPS. There are neither base stations nor infrastructure support to coordinate the activities of subsets of nodes. Therefore, all the nodes have to collectively make decisions. We assume that the nodes are stationary. This assumption about node mobility is typical for sensor networks. All sensors transmit at the same power level and hence have the same transmission range ($T_r$).

All communication is over a single shared wireless channel. A wireless link can be established between a pair of nodes only if they are within wireless range of each other. The *KOCA* algorithm only considers bidirectional links. It is assumed the MAC layer will mask unidirectional links and pass bidirectional links to *KOCA*. We refer to any two nodes that have a wireless link as 1-hop or immediate neighbors. Nodes can identify neighbors using beacons. Unless mentioned otherwise, we assume that the communication environment is contention-free and error-free; hence, sensors do not have to retransmit any data. The Multiple Access with Collision Avoidance (MACA) protocol [43] may be used to allow asynchronous communication while avoiding collisions. Other MAC protocols such as TDMA [44] may be used to provide collision-free MAC layer communication. We study the effect of contention on the performance of *KOCA* in Section V.

### B. Definitions and Notations

A sensor network can be modeled as a graph $G = (V, E)$, where vertices in the graph represent sensor nodes and two vertices are connected by an edge if the corresponding sensor nodes can communicate with each other.

We use the following notations and definitions:

- *Network size* ($n$): the number of sensor nodes in the network, $n = |V|$. Changing the network size changes the node density ($\mu = n/l^2$) since the area is fixed, where $l$ is the side length of the square deployment area.
- *Cluster radius* ($k$): the maximum distance between any node in the cluster and the cluster head, where the distance between any two nodes is the minimum number of hops between them.

[1] A homogeneous sensor network is commonly used in literature. Using a heterogeneous network, of both sensors and more powerful base stations can be handled in the same way as described in this paper. For example, only base stations can be allowed to become cluster heads, while sensors are prevented from volunteering as cluster heads. In this case, the number of base stations will be a design parameter, that can be selected by techniques similar to the ones described in this paper.

- *Average Node Degree ($d$):* the average node degree in the network. The node degree of a node $u$, is the number of nodes that are neighbors of $u$. The relation between the average node degree ($d$) and the transmission range ($T_r$) of a node is given by [45]:

$$d = \frac{n\pi T_r^2}{l^2} = \mu\pi T_r^2 \tag{1}$$

- The cluster head probability ($p$): The probability that a node will be a cluster head. The average number of clusters is $pn$. Therefore, increasing $p$ will increase the number of clusters in the network.
- *Closed $k$-Neighbor Set* of a node $u$ ($N_k[u]$): is the set of nodes that are reachable to a node $u$ in at most $k$ hops, including $u$ itself.
- *$k$-Dominating Set* (*KDS*) ($S$): is defined as a subset of $V$ such that each vertex in $V - S$ is within distance $k$ from at least one vertex in $S$, where $k \geqslant 1$ is an integer.
- Overlapping degree between two clusters $A$ and $B$: is the number of common nodes between the two clusters.
- Induced Overlapping Graph ($\langle S \rangle$): is a graph whose vertices $\in S$ and there is an edge between nodes $u$ and $v \in S \iff |N_k[u] \cap N_k[v]| \geq 1$

### C. Problem Formulation

Given a graph, $G = (V, E)$, representing a sensor network, we formulate the overlapping clustering problem as finding the set of cluster-head nodes $S$ such that the following three conditions are satisfied:

1) *Coverage Condition*: $S$ is a *KDS*. This means that each node is either a cluster head or within $k$ hops from a cluster head (i.e. $N_k[S] = V$).
2) *Overlapping Condition*: For each node $u \in S \, \exists$ at least one node $v \in S$ such that $|N_k[u] \cap N_k[v]| \geq o$, where $o$ is a certain threshold. In other worlds, for each cluster $A$, there exists at least one other cluster $B$ that overlaps with it with overlapping degree $\geq o$.
3) *Connectivity Condition*: $\langle S \rangle$ is connected.

One way of approaching the problem is to find the minimum *KDS* (*MKDS*). This is a desirable goal as it can help in decreasing the control overhead by restricting the broadcast of route discovery and topology update messages to a small subset of nodes. However, there is no known efficient centralized algorithm for obtaining an exact solution to the *MKDS* problem and the corresponding decision problem is NP-hard [46], even for the special simplified case of unit-disk graphs, which are common in sensor networks. Further aspects of the computability of *MKDS* are discussed in [46] and [47].

In the next section, we present the *KOCA* clustering algorithm. *KOCA* is a distributed randomized heuristic algorithm that achieves the above three conditions efficiently.

## III. The *KOCA* Heuristic

There are three types of nodes in the clusters generated by *KOCA*: Cluster heads (CHs), boundary nodes (BNs), and normal nodes. A cluster head maintains a graph representing its cluster along with information about adjacent clusters and how to reach them. A boundary node belongs to more than one cluster and may work as a gateway between these clusters as needed. Normal nodes are internal nodes that belong only to one cluster. In the balance of this section, we discuss the necessary data structures maintained at each node, followed by an example to illustrate the clusters generated by *KOCA*. We then describe the cluster head selection process and describe cluster membership.

### A. Data Structures

Each node maintains the following variables:

- Node ID (*NID*): A unique ID assigned to each node before deploying the network.
- Adjacent Clusters Table (*AC_table*): A table maintained by CH nodes to store information about adjacent clusters. The table consists of tuples in the form (*CHID*, *BNL*), where *CHID* is the CH node ID, and *BNL* is a list of *boundary node* IDs (Fig. 1).
- Boundary Table (*CH_table*): A table maintained by each node to store information about the clusters known to this node. If the table contains more than one entry, this means that the node is a *boundary node*, otherwise, the node is a normal node. The table consists of tuples of the form (*CHID*, *HC*, *prev*), where *CHID* is the CH node ID, *HC* is the number of hops leading to this cluster head, and *prev* is the node ID of a 1-hop neighbor node that can lead to this CH node using minimum number of hops (part of the shortest path).

### B. Example

Fig. 1 gives an example for the output of the *KOCA* algorithm. In the figure, there are three clusters with three corresponding cluster heads $A, B, C$. The figure shows an example of the adjacent clusters table maintained at the head of a cluster $C$. Since Cluster $C$ overlaps with two clusters, Node $C$'s table contains 2 entries. For example, the first entry indicates that Cluster $C$ can reach Cluster $A$ by using the boundary nodes $B_1$ or $B_2$.

The figure also shows an example of the boundary table at boundary node $B_4$. Since boundary node $B_4$ belongs to two clusters, its table contains two entries. For example, the first entry in the table indicates that $B_4$ belongs to Cluster $C$ and can reach its cluster head in two hops through node $S_1$.

### C. Cluster Head Selection

The essential operation in any clustering protocol is to select a set of cluster heads among the nodes in the network, and group the remaining nodes around these heads. *KOCA* does this in a distributed fashion, where nodes make autonomous decisions without any centralized control. The algorithm initially assumes that each sensor in the network becomes a cluster head with probability $p$. Each cluster head then advertises itself as a cluster head to the sensors within its radio range. This advertisement is forwarded to all sensors that are no more than $k$ hops away from the CH through controlled flooding. The advertisement ($CH_AD$) message's header include *SID*, *CHID*, and *HC*; where *SID* is the sender node ID, *CHID* is cluster head ID, and *HC* is the number of hops leading to the *CH* node. The *SID* field is used to update the $CH_table$.prev field such that each node knows the path to the cluster head. The *HC* field is used to limit the flooding of the $CH_AD$ message to $k$ hops. As we explain later, a sensor that receives such advertisements joins the cluster even if it already belongs to another cluster. Since the advertisement forwarding is limited to $k$ hops, if a sensor does not receive a *CH* advertisement within a reasonable time duration, it can infer that it is not within k hops of any cluster head and hence become a *CH*. In *KOCA*, the maximum time that a node should wait for *CH* advertisement messages is set to $t(k) + \delta$, where $t(k)$ is the time needed for a message to travel $k$ hops and $\delta$ is the maximum time needed for any node to finish bootstrapping and start the clustering process.

### D. Cluster Membership

Each node maintains a table, $CH_table$, that stores information about the clusters it belongs to. Upon receiving a new $CH_AD$ message, a node will add an entry in its *CH*_table. In case a similar message was received from the same cluster, the node will check the hop count, i.e. the *HC* field in the recent message, and will then update *HC* and *prev* fields in the corresponding entry in the *CH_table* if the recent message came over a shorter path. Often a message traveling the shortest path in terms of the number of hops would arrive first. However, delay may be suffered at the MAC or link layers. If $CH_table$ contains more than one entry, this means that the node is a boundary node. For every entry in its $CH_table$, a node sends a join request (*JREQ*) message to the CH in order become a member of the corresponding cluster. To limit the flooding, the message is unicast using the field $CH_table$.prev. The *JREQ* message has the form *[JREQ, RID, SID, CHID, nc,* $(CHID)_{0..nc}$*]* where RID is the receiver node ID (i.e. $CH_table$.prev), SID is the ID of the node that will join the cluster, CHID is the ID of the *CH* node responsible for this cluster, $nc$ is the number of clusters that this node can hear from ($|CH_table|$), and $(\text{CHID})_{0..nc}$ are 0 or more clusters that this node can hear from. Each cluster head maintains a list of all cluster members, a list of adjacent clusters, and a list of boundary nodes to reach those clusters along with the maximum hop count to reach the adjacent cluster. There can be multiple boundary nodes between overlapping clusters. Moreover, a node can be a boundary node for more than two overlapping clusters. The *CH* node also will enforce a time-out for *JREQ* which is set in *KOCA* to $ct(k) + \delta$ ; where $c$ is a constant that depends on the MAC protocol, node density and the value of $p$.

The events of the *KOCA* clustering algorithm are listed in Table I. A finite state machine for the protocol is given in Fig. 2. The activities of the *KOCA* clustering algorithm are given in Appendix A along with the proof of the correctness of the algorithm.

Note that *KOCA* terminates in $O(k)$ steps. Typically $k$ is a constant, so the clustering process terminates in a constant number of iterations regardless of the network size.

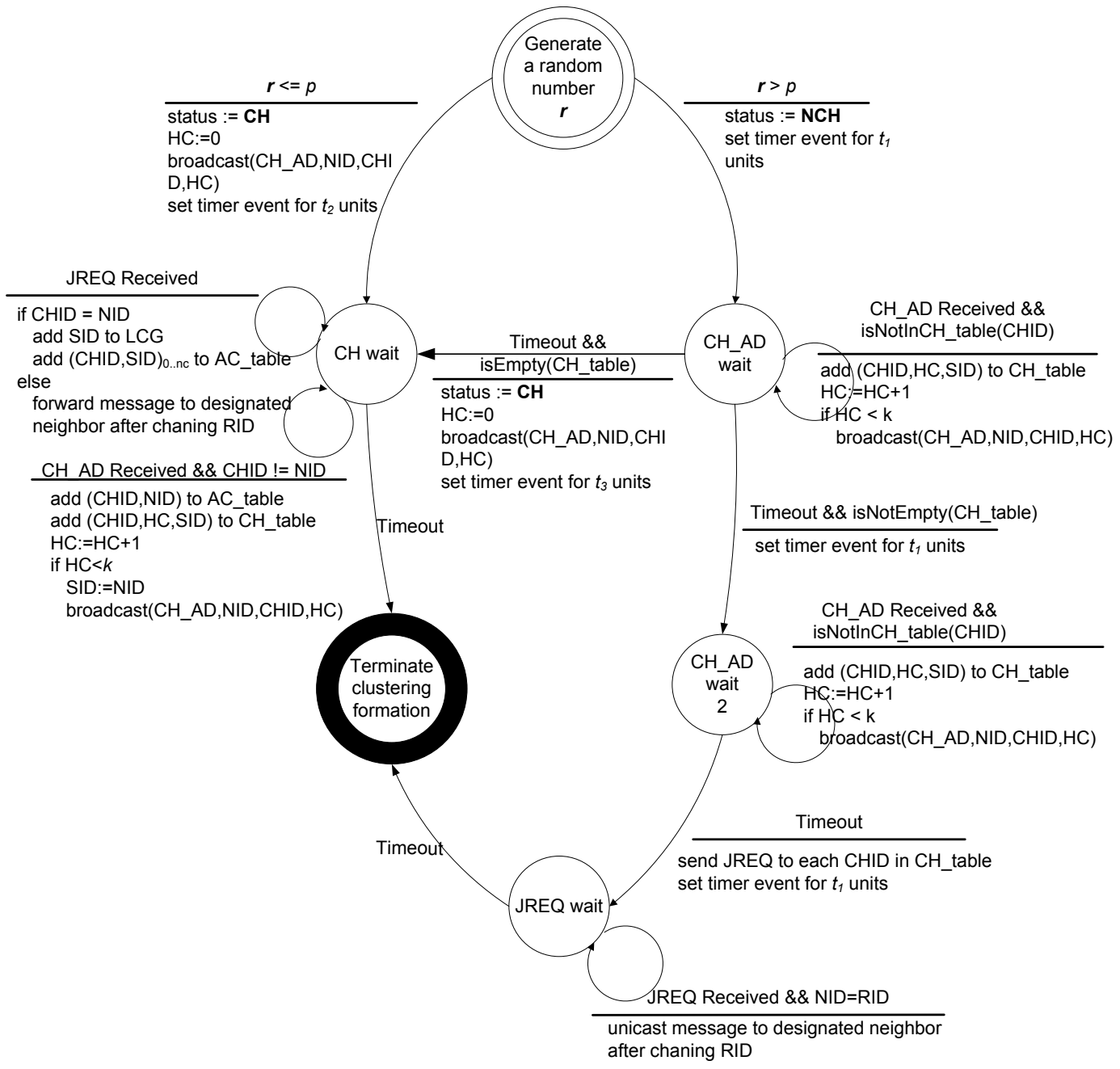

Fig. 2. The finite state machine of the *KOCA* algorithm.

## IV. Analysis

In this section, we study analytically the effect of different parameters of *KOCA* on the clustering process. These results are validated by simulation in Section V. We use the following assumptions in the analysis:

- Sensors are deployed uniformly in a square area with side length of $l$.
- The largest radius of a cluster, $R$, can be approximated by a circle with radius $R = kT_r$.
- The cluster head is located at the center of the cluster.

The second assumption leads to an upper bound of the different quantities analyzed in this section. The third assumption is reasonable due to the nature of the *KOCA* algorithm where the cluster formation starts from the cluster head and propagates outward towards the edges of the cluster (Fig. 4).

### A. Average Number of Nodes per Cluster

We start by deriving an upper bound on the average number of nodes per cluster (cluster size). Let $N_c$ be a discrete random variable representing the cluster size. $N_c$ can be expressed by the following binomial distribution:

$$P(N_c = m) = P_c^m (1 - P_c)^{n-m} \begin{pmatrix} n \\ m \end{pmatrix} \tag{2}$$

where $n$ is the network size and $P_c$ is the probability that a node is inside the circle representing the cluster. $P_c$ can be calculated as:

$$P_c = \frac{\pi R^2}{l^2} = \frac{\pi k^2 T_r^2}{l^2} \tag{3}$$

Substituting from equation 1 we get

$$P_c = \frac{dk^2}{n} \tag{4}$$

| Event Name | Description |
|---|---|
| Initialization() | An event executed once to initialize the status of the node. |
| CH_AD_Received (*SID, CHID, HC*) | An event triggered when CH_AD message is received. |
| JREQ_Received (*RID, SID, CHID, nd, (NID, RSID,NID)*$_{1..nd}$, *nc, (CHID)*$_{0..nc}$) | An event triggered when JREQ message is received. |
| ChangeStatus | An event triggered when the CH_AD_WAIT timer fires indicating that an NCH node should either change its status to CH node or join a cluster if any. |
| EndClusterFormationPhase | An event triggered when the JREQ_WAIT timer fires indicating that a CH node should terminate the clustering phase and start the Local Location Discovery (LLD) phase. |

TABLE I
EVENTS SUMMARY OF THE *KOCA* CLUSTERING ALGORITHM

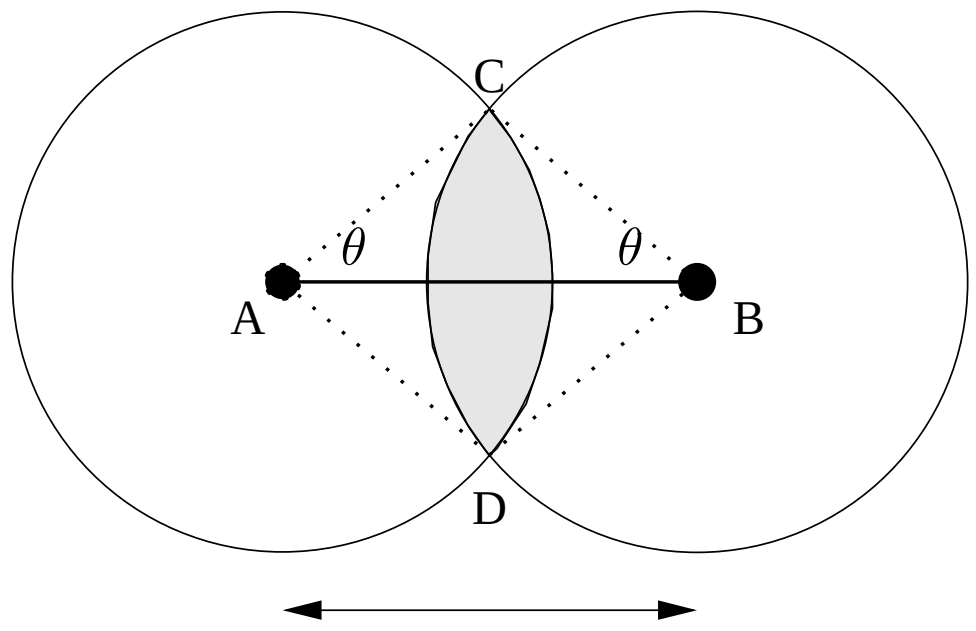

Fig. 3. Overlapping Degree (***O***) between two overlapping clusters defines the number of nodes in the intersection of these two clusters.

Hence, the average cluster size ($E(N_c)$) is:

$$E(N_c) = nP_c = dk^2 \tag{5}$$

The above equation shows that the average cluster size is linearly proportional to the average node degree ($d$) and quadratically proportional to the cluster radius ($k$).

### B. Average Overlapping Degree

We calculate an upper bound on the average overlapping degree (AOD). Assume that $A$ and $B$ are any two cluster head nodes. Let the overlapping degree between the two clusters (***O***) be a random variable, where ***O*** = $|N_k[A] \cap N_k[B]|$ and $N_k[A] \cap N_k[B] \neq \phi$. Notice that the overlapping degree is defined only for overlapping clusters (i.e. the random variable ***O*** does not take the value 0). We define *AOD* as the mean of this random variable ***O***.

As shown in Fig. 3, the two clusters $A$ and $B$ are represented by two symmetric circles of radius $R = kT_r$, where $T_r$ is the sensor's transmission range. We start by calculating the expected area of intersection between the two clusters. Let $W$ be the euclidian distance between the two CH nodes. Then, $W$ is a continuous random variable that can take values ranging from $0$ to $2R$. The two clusters are completely overlapping if $W = 0$ and there is no overlapping if the distance between the two cluster heads is greater than or equal $2R$. Let $F(w)$ and $f(w)$ be the CDF and PDF of the random variable $W$ respectively.

$$F(w) = P(W < w) = \frac{\pi w^2}{\pi (2R)^2} = \frac{w^2}{4R^2} \tag{6}$$

and

$$f(w) = \frac{dF(w)}{dw} = \frac{w}{2R^2} \tag{7}$$

To obtain the average intersection area between two symmetric circles $A$ and $B$ ($E(I)$), assume that the two circles intersect in some area $I_{AB}$. Let $r$ be the radius length and $w$ be the distance between the two centers $A$ and $B$ as shown in Fig. 3. then the intersection ($I_{AB}$) can be calculated as follows:

$I_{AB}$ = 2 (area of sector $CBD$ - area of triangle $CBD$)
Area of sector CBD = $\frac{1}{2}.2\theta.R^2 = \theta.R^2$
$\therefore I_{AB} = 2(\theta R^2 - \frac{1}{2}.R^2 \sin 2\theta) = (2\theta - \sin 2\theta)R^2$
where $w = 2R\cos\theta$ (using cosine rule).

Using these results, the average overlapping degree can be calculated as:

$$AOD = E(\boldsymbol{O}) = E(\mathbf{I})\mu = \frac{dk^2}{4} \tag{8}$$

The above equation shows that the average overlapping degree is linearly proportional to the average node degree $d$ and quadratically proportional to the cluster radius $k$.

### C. Overall Communication Overhead

In this section, we calculate an upper bound of the average number of messages transmitted by a node, the overall communication overhead per cluster, and the overall communication overhead for the network. Recall that there are two phases in the *KOCA* protocol: the cluster head advertisement phase (CHAD phase) and the join request phase (JREQ phase). The overall communication overhead is the sum of the number of messages in these two phases. We start by the total number of messages per cluster. Using Eq. 5

$$E(N_c) = nP_c = dk^2 \triangleq E_k(N_c) \tag{9}$$

the average number of nodes that are exactly $k$ hops away from the cluster head ($n_k$) is:

$$n_k = E_k(N_c) - E_{k-1}(N_c) = dk^2 - d(k-1)^2 = d(2k-1) \tag{10}$$

Using the above results, we can calculate the average number of CH_AD messages sent during the CHAD phase. The CH_AD messages are forwarded through the edges of a spanning tree, rooted at the CH, of the cluster graph as shown in Fig. 4.

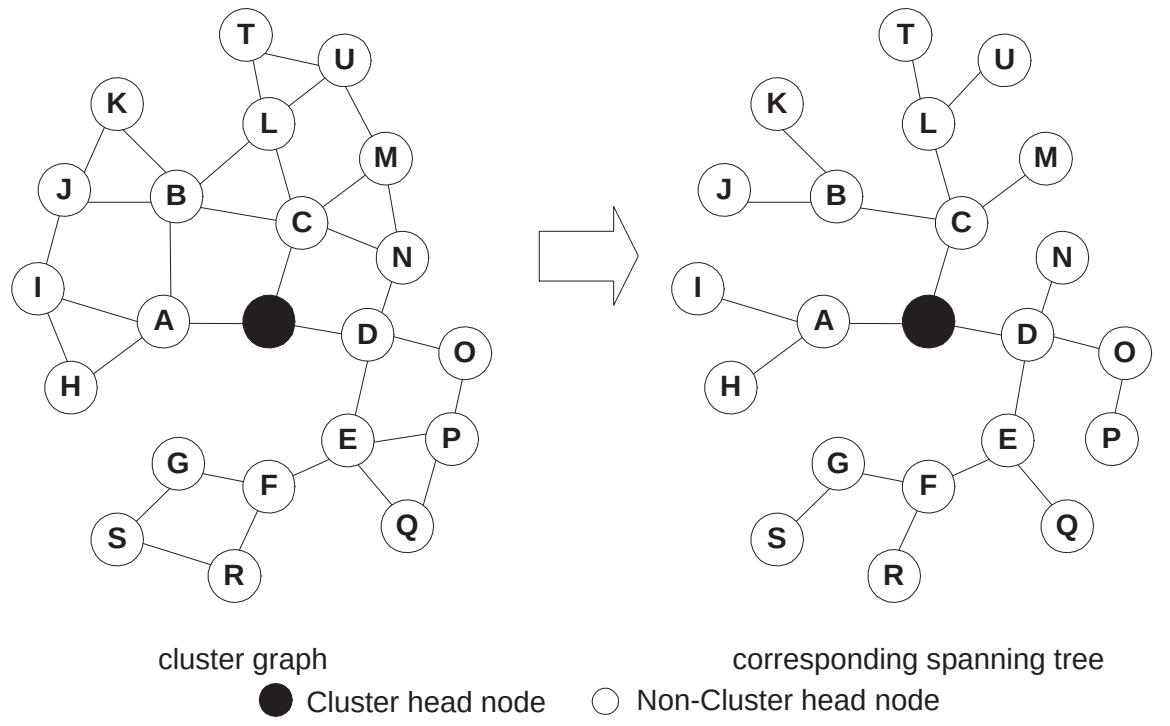

Fig. 4. The CH_AD message will follow a spanning tree rooted at the CH node ($k = 5$)

Let $M_{CHAD}$ be the average number of CH_AD messages broadcasted within the cluster. Then $M_{CHAD}$ is equal to the average number of non-leaf nodes in breadth-first tree of the graph rooted at the CH node.

$$M_{CHAD} = 1 + \sum_{i=1}^{k-1} n_i \tag{11}$$

where $n_i$ is the expected number of nodes that are exactly $i$ hops way from the CH node (Eq. 10). Substituting from Eq. 10 and simplifying the expression, we reach the following:

$$M_{CHAD} = 1 + \frac{2d(k-1)^2}{2} = O(dk^2) \tag{12}$$

Using a similar approach, we can calculate the average number of JREQ messages ($M_{JREQ}$) *unicast* from non-CH nodes to the CH node. We assume that we do not do any aggregation of the messages. If message aggregation is used, the overall communication overhead will improve. So the above analysis is a worst case analysis. A JREQ message, unicast from a leaf node in the spanning tree, will be forwarded $k$ times till it reach the CH node. Therefore, $M_{JREQ}$ can be calculated as follows:

$$M_{JREQ} = kn_k + (k-1)n_{k-1} + \ldots + 2n_2 + n_1 = \sum_{i=1}^{k} in_i \tag{13}$$

Substituting from Eq. 10 and simplifying the expression, we reach the following expression:

$$M_{JREQ} = \frac{dk(4k-1)(k+1)}{6} = O(dk^3) \tag{14}$$

Therefore, the overall communication overhead per cluster is $M_{Cluster} = M_{CHAD} + M_{JREQ} = O(dk^3)$. Therefore, the overall communication overhead for the network is $M_{Network} = M_{Cluster}np = O(ndpk^3)$ and the communication overhead per node is $M_{Node} = M_{Network}/n = dpk^3$, which is independent from the network size.

### D. Summary

In this section, we showed analytically the following effect of the parameters, $d, p$, and $k$, on the clustering process:

- The average number of nodes per cluster ($N_c$) is linear in $d$ and quadratic in $k$.
- The cluster head probability ($p$) does not affect the average number of nodes per cluster nor the average overlapping degree (*AOD*).
- The average overlapping degree (AOD) is linearly proportional to the average node degree ($d$) and quadratically proportional to the cluster radius ($k$).
- The communication overhead per node is linearly proportional with $d$ and $p$ and cubically proportional with $k$.
- The communication overhead per node is independent from the network size ($n$).

## V. Performance Evaluation

This section studies the performance of the *KOCA* algorithm and the effect of different parameters on the clustering process. Our goals are: (1) to show that with proper selection of the input parameters ($p$, $k$, $d$), the proposed clustering algorithm meets the conditions listed in Section II, (2) to validate the analytical results derived above, and (3) to show that KOCA is scalable in terms of communication overhead.

### A. Simulation Environment

The *KOCA* clustering algorithm was implemented using NS2 simulator refns2. Nodes are spread uniformly over a square area of 100×100 unit area. All experiments were performed over 150 different topologies representing different network sizes ($n$) ranging from 50 to 800 sensor nodes. For each topology, the transmission range of each node ($T_r$) was varied in order to achieve different average *node degree* ($d$) ranging from 7 to 21. In a sensor network with a uniform distribution of nodes, in order to guarantee network connectivity, the average node degree should be at least 6 [48]. Hence, we chose the minimum average node degree to be 7. This is a reasonable value for sensor networks which typically have dense deployment. The cluster radius ($k$) ranges from 1 to 5. The cluster head probability ($p$) was varied from 0.05 to 0.5. We repeat the experiment 30 times for each topology. For all experiments, the simulation results stay within 2-6% of the sample mean with 95% confidence level. Details of the error and confidence interval curves can be found in the accompanying technical report [49]. Except for Section V-H, we use a contention-free *TDMA* MAC protocol and an error-free environment.

### B. Performance Metrics

We use the following performance metrics:

1) *Percentage of Covered Nodes (CN)*: this metric captures whether the generated clusters satisfy the *coverage condition* defined in Section II or not. *CN* is defined as the percentage of nodes that are either cluster heads or within $k$-hops from a cluster head after the first wave of CH advertisement is propagated though the network.
2) *The Average Overlapping Degree (AOD)*: this metric reflects whether the generated clusters satisfy the *overlapping condition* defined in Section II or not. *AOD* is defined as the average overlapping degree between any two overlapping clusters in the network.

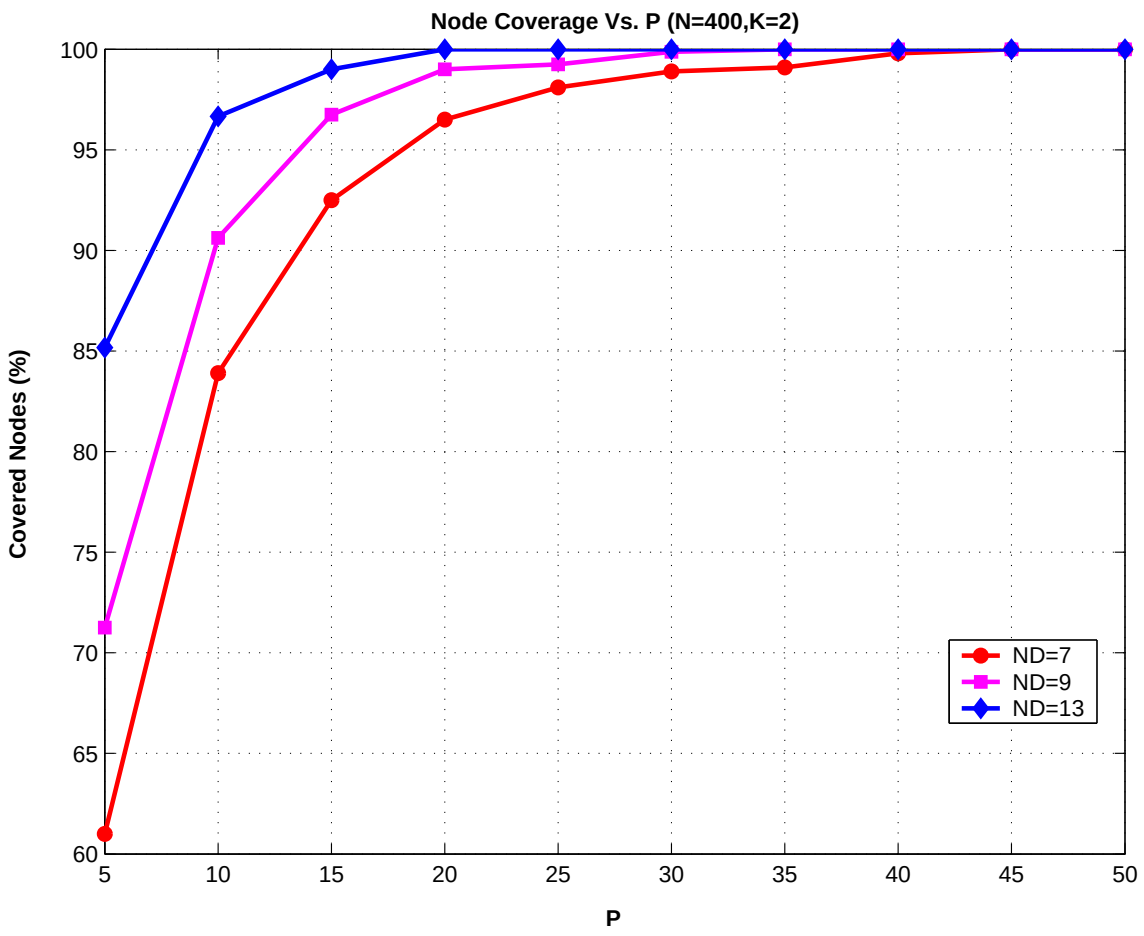

(a) Effect of changing $p$ for different $d$

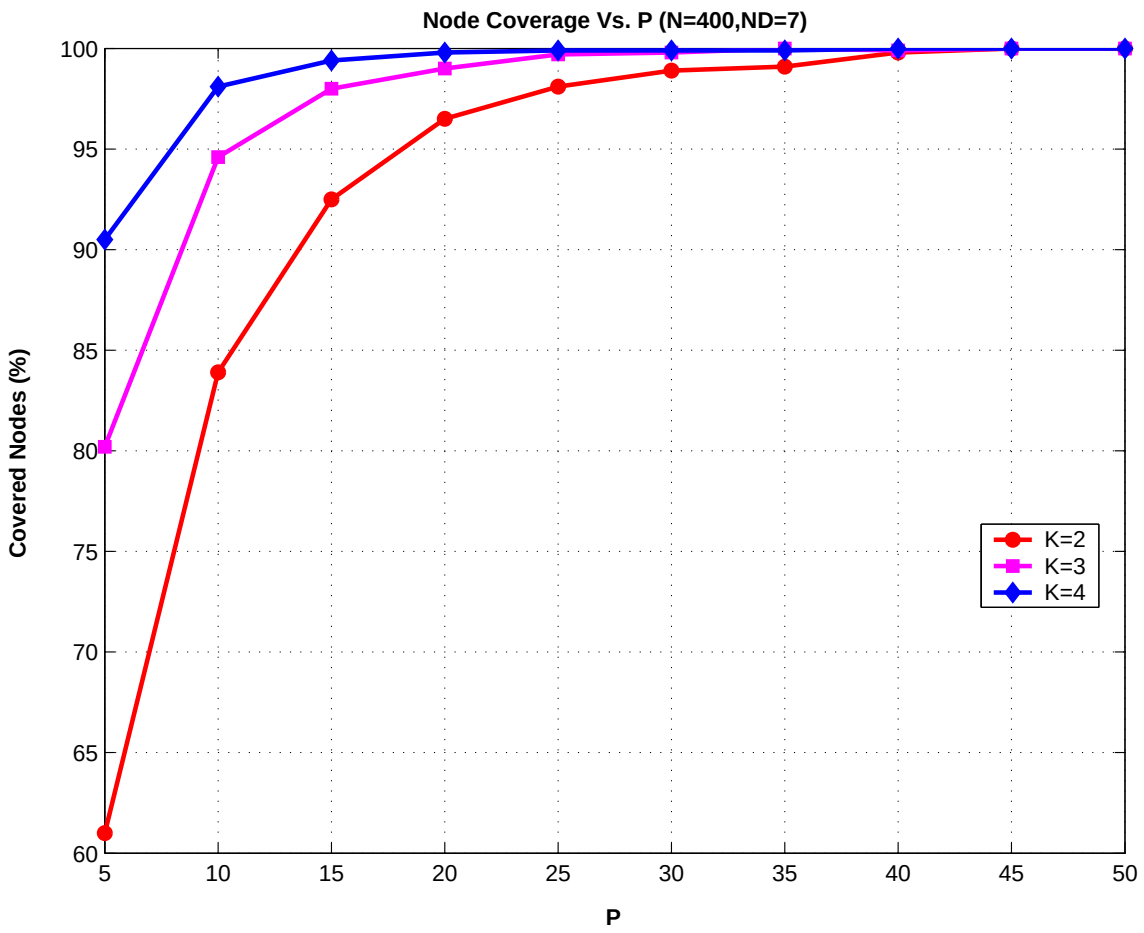

(b) Effect of changing $p$ for different $k$

Fig. 5. The relation between cluster head prob. ($p$) and percentage of covered nodes.

3) *Normalized Standard Deviation of Overlapping Degree*: this metric reflects how uniform the overlapping degree is between different clusters. This metric quantifies how close the average overlapping degree is to the minimum overlapping degree. This is important for some applications as described in Section V-I.
4) *The Connectivity Ratio* (*CR*): this metric reflects whether the generated clusters satisfy the *connectivity condition* defined in Section II or not. The connectivity ratio (*CR*) is defined as ratio between the number of nodes in the largest spanning tree of the induced overlapping graph ($\langle S \rangle$) to the number of *CH* nodes ($|S|$). If *CR* = 1, this means that $\langle S \rangle$ is a connected graph.
5) *The Average Cluster Size* ($N_c$): the average number of nodes per cluster.
6) *Communication Overhead*: this metric reflects the total number of messages transmitted in the network.

### C. Effect of the Parameters on Coverage

Fig. 5 shows the effect of changing the cluster head probability ($p$) on the percentage of covered nodes (*CN*) for different values of $d$ and $k$. We can see from the figure that increasing $p$ increases the coverage. We can also see from the figure that for each combination of ($k$, $d$), there is a value for $p$ that guarantees 100% coverage. The figures also show that the curves saturate around $p = 0.15$. We use this value for the rest of this section.

From the same figure, we notice that increasing the average node degree ($d$) or cluster radius ($k$) leads to increasing the percentage of covered nodes. Therefore, to achieve a high probability of coverage for a fixed $p$, we need to tune the values of $k$ and $d$. Since for a fixed network, increasing $d$ can only be achieved through increasing the transmission range, it affects nodes energy consumption significantly. Therefore, it is recommended that the desired coverage is obtained through changing the value of $k$.

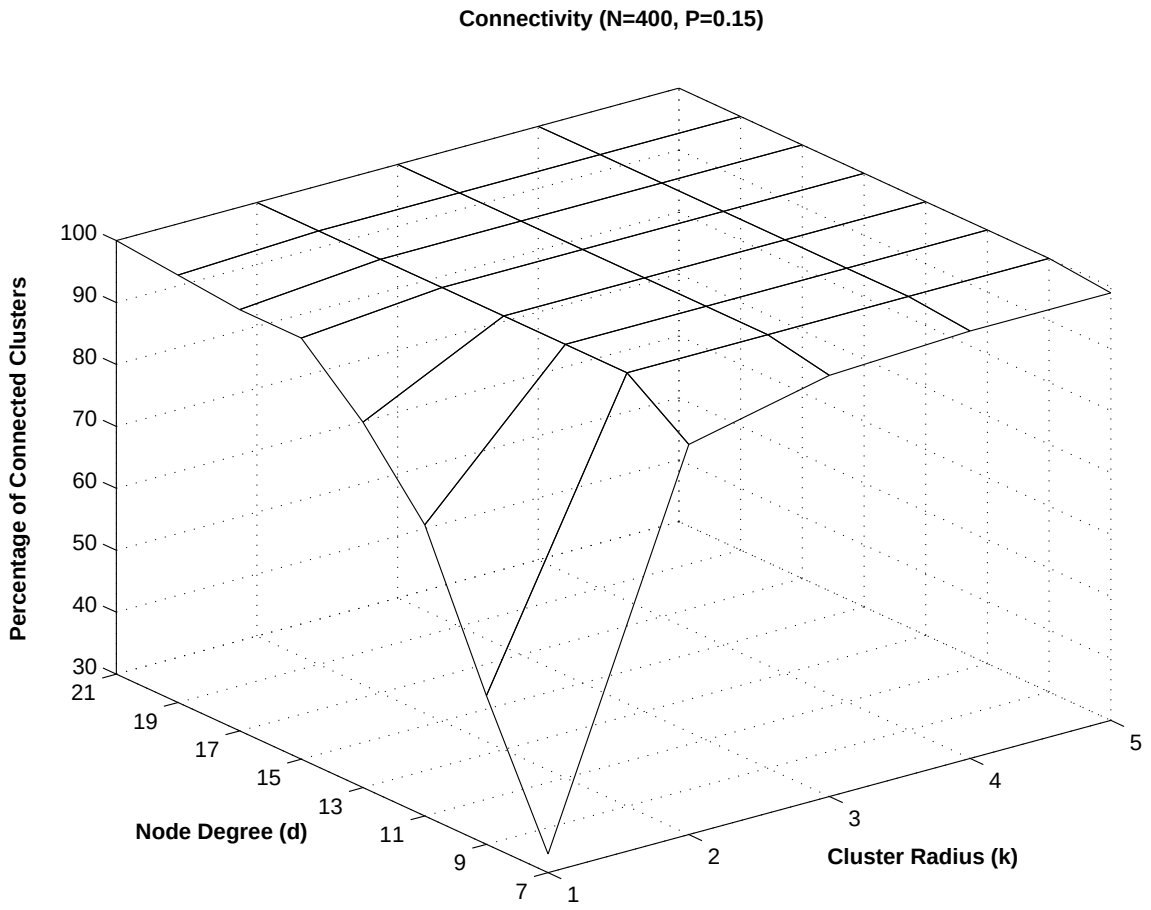

Fig. 6. The effect of the average node degree ($d$) and number of hops ($k$) on the connectivity ratio (CR).

### D. Effect of the Parameters on Connectivity

Fig. 6 shows the effect of changing $k$ and $d$ on the connectivity ratio for $p = 0.15$. The figure shows that connectivity increases as $d$ or $k$ increases. Moreover, we can achieve 100% connectivity. This means that for any cluster head, there is a path of less than $2k$ hops to at least another cluster head (i.e. there is at least one boundary node between the two clusters).

### E. Effect of the Parameters on Overlapping Degree

This section studies the effect of different parameters on the average overlapping degree between clusters. Fig. 7(a) shows an interesting anomaly. Although one may think that increasing $p$ (i.e. increasing the number of cluster heads and hence clusters) should increase the average overlapping degree (AOD), the results shows that $p$ has no effect on AOD regardless of the values of the other parameters ($d$, $k$) and network size ($n$). The intuition is that there are two opposing factors: (1) as $p$ increases, the number of clusters increases and the overlapping between clusters increases, (2) however, the number of pairwise intersections between clusters increases too. Since the *AOD* is the ratio between these two quantities, and they change with the same rate as proved analytically, the *AOD* is independent of $p$.

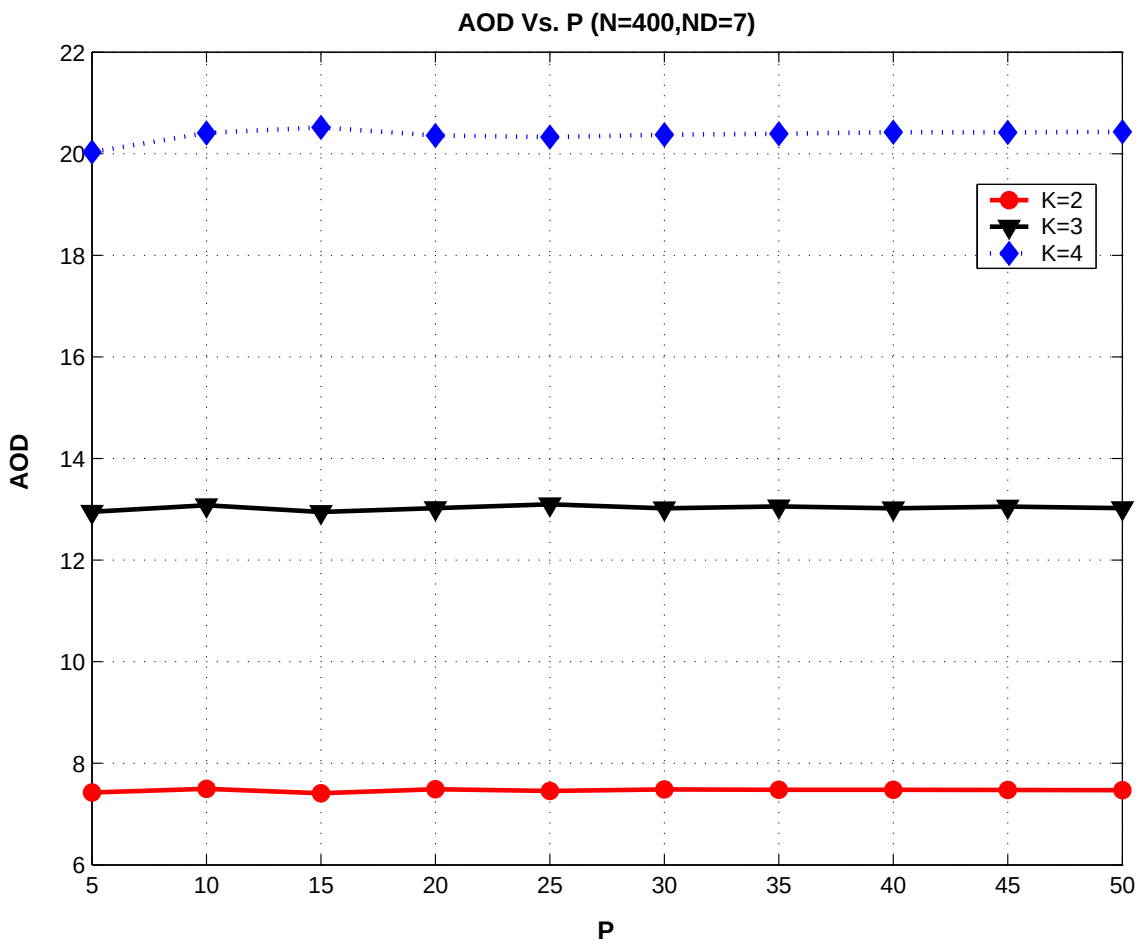

(a) $p$ has no effect

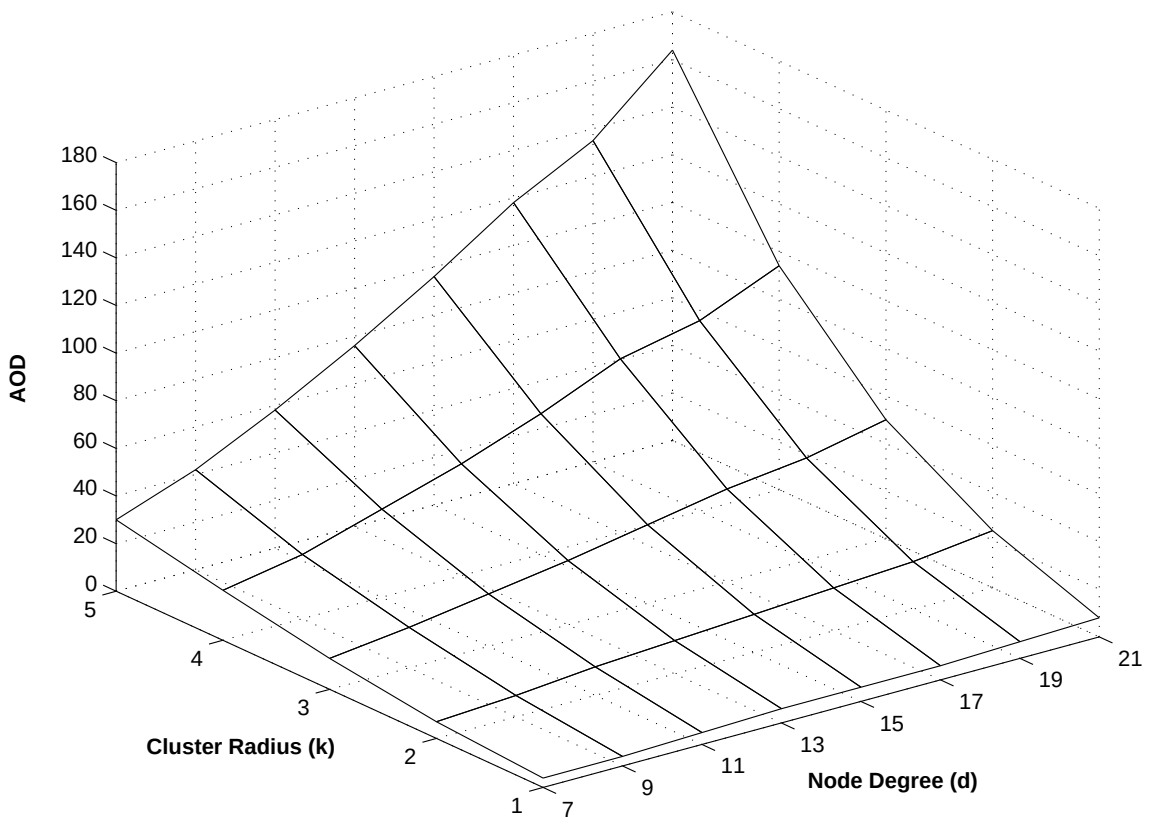

(b) The impact of average node degree $(d)$ and cluster radius $(k)$

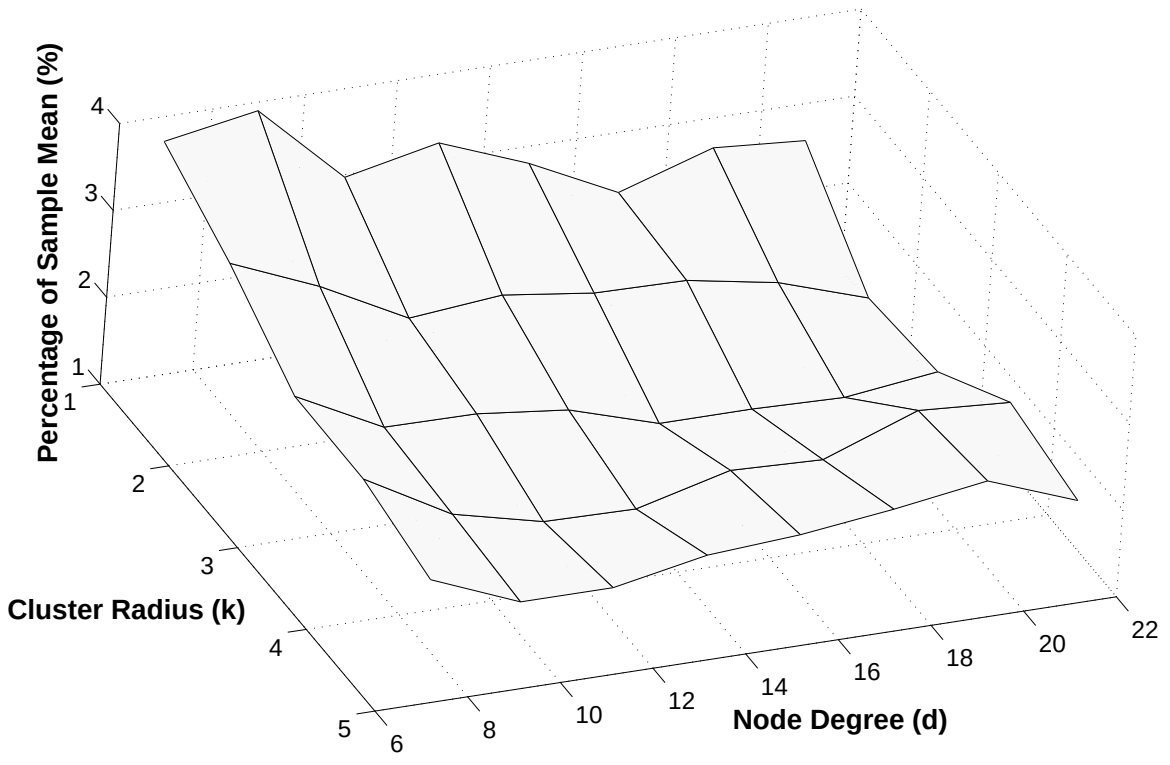

(c) Stdev of overlapping degree

Fig. 7. The impact of different parameters on average overlapping degree (AOD).

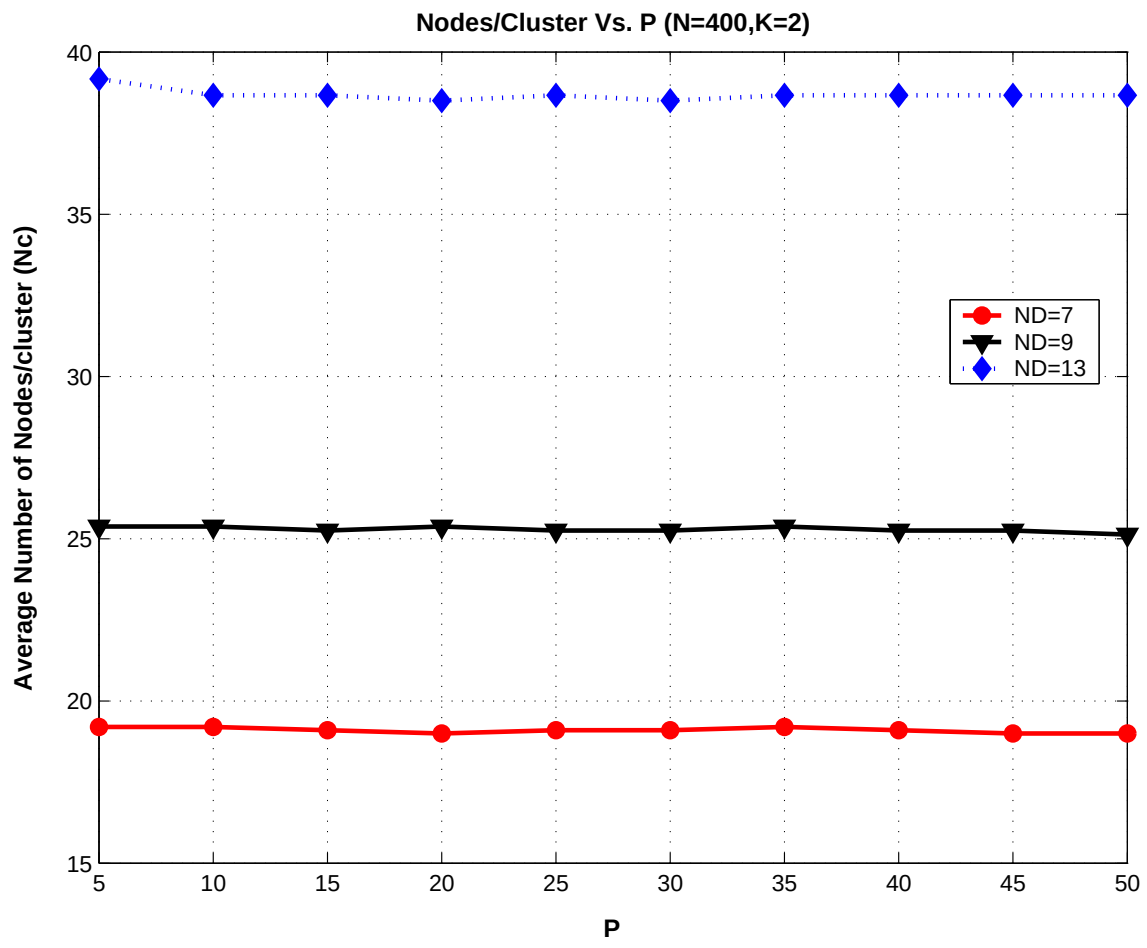

(a) The cluster head prob. $(p)$ has no effect on the average number of nodes/cluster

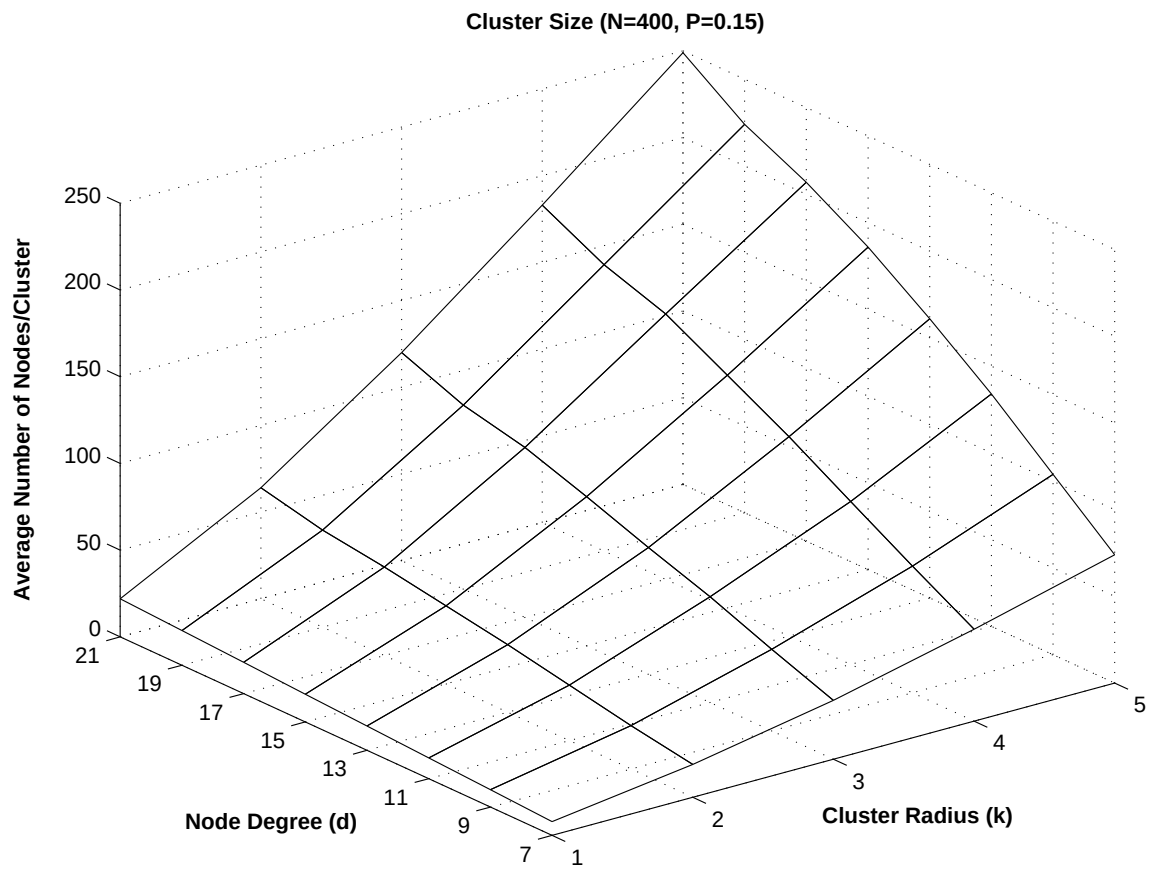

(b) The effect of average node degree $(d)$ and cluster radius $(k)$ on number of nodes/cluster $(N_c)$

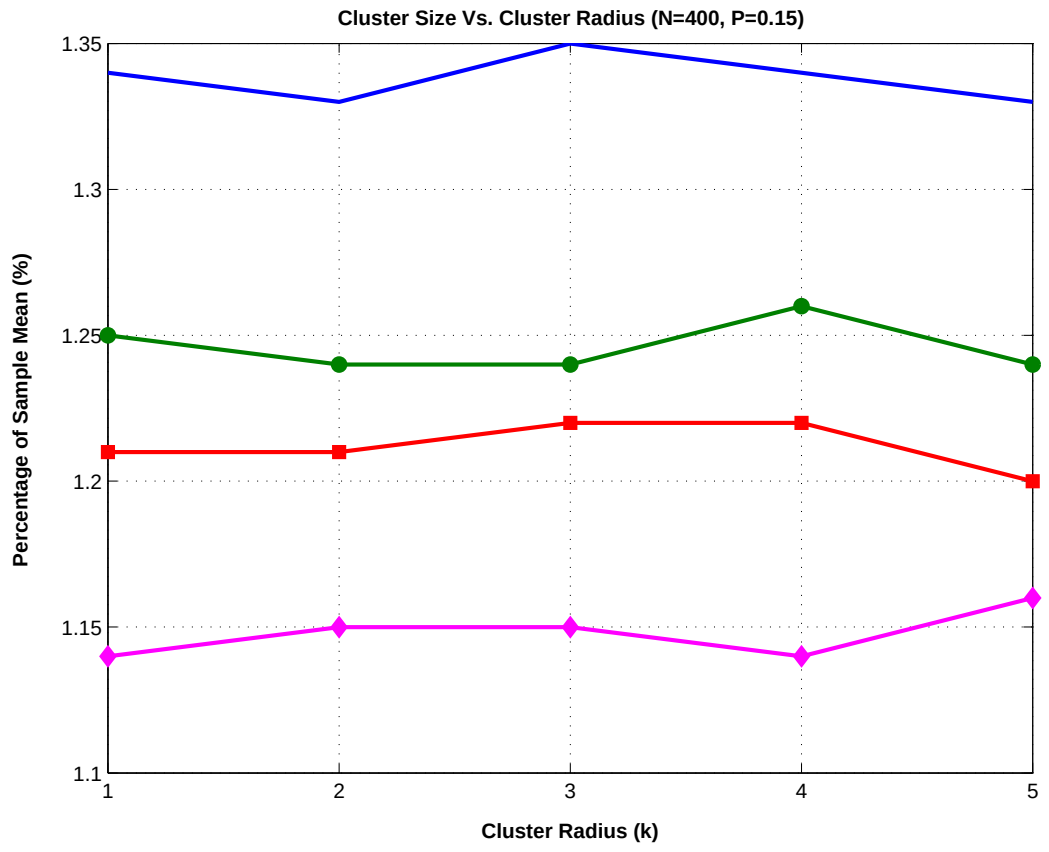

(c) Normalized standard deviation of the cluster size as $k$ and $d$ change

Fig. 8. The effect of different parameters on cluster size metrics.

This leaves us with two parameters to control the overlapping between clusters: $d$ and $k$. As shown in Fig. 7(b), the *AOD* is linearly proportional with $d$. Notice that *AOD* can never exceed the network size $n$ so the curve saturates at $n$. On the other hand, increasing the cluster radius $(k)$ will increase the *AOD* quadratically as shown in Fig. 7(b). This confirms the analytical results in Section IV-B.

For many applications, having an *AOD* of 10 between clusters is more than enough. For example in localization, an *AOD* of 3 is sufficient for locating nodes in 2D. Similarly, in routing protocols having 10 gateway nodes between clusters is more than enough [29], [30]. It is clear that we can guarantee an *AOD* of more than 10 using small $d$ (i.e. low transmission range) and small cluster radius $(k = 2)$.

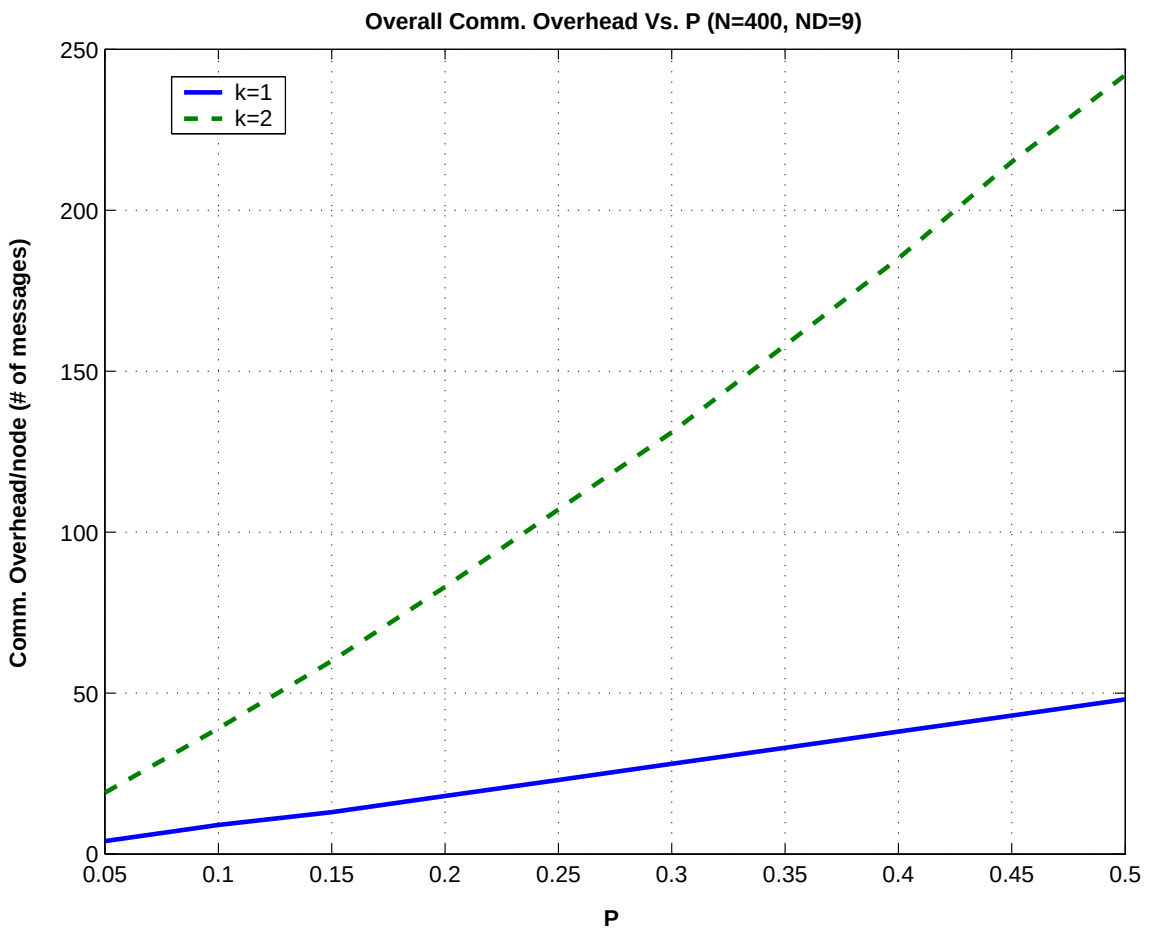

(a) Effect of cluster head prob. ($p$)

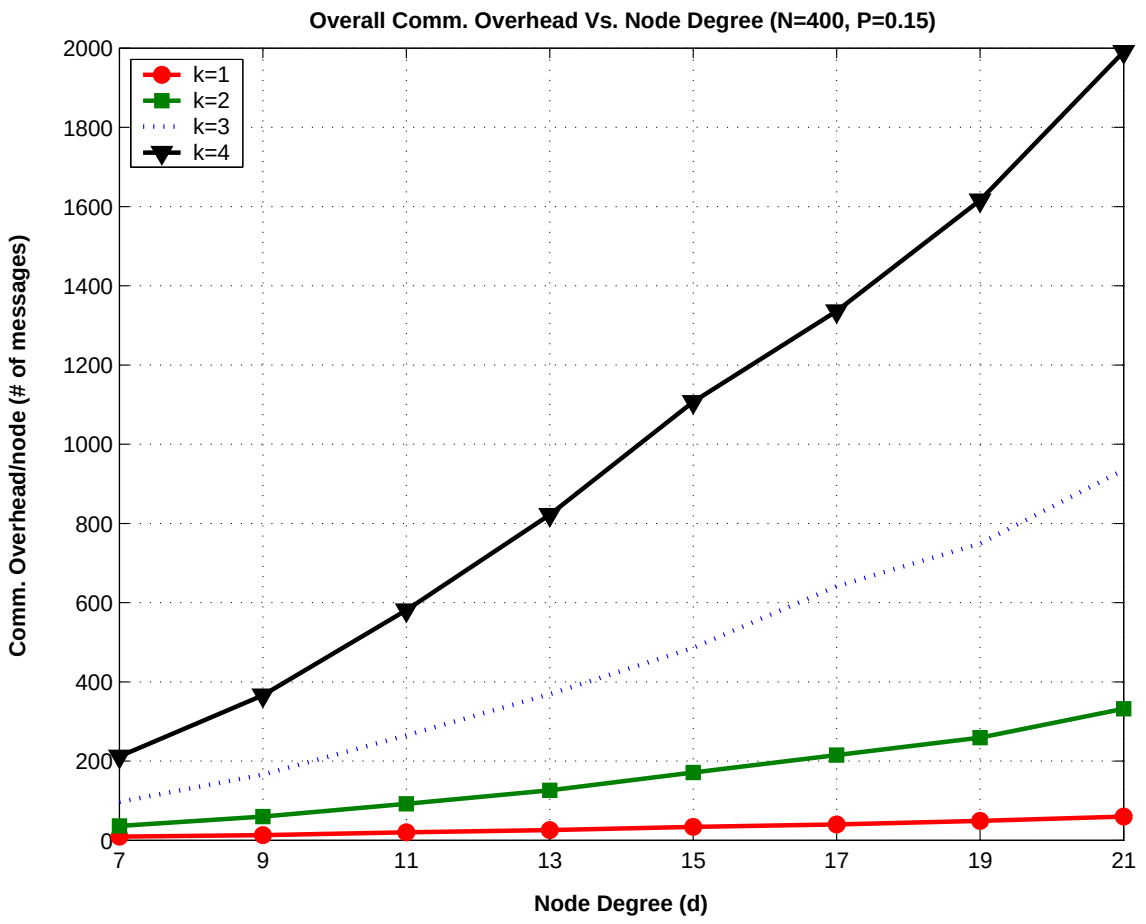

(b) Effect of average node degree ($d$)

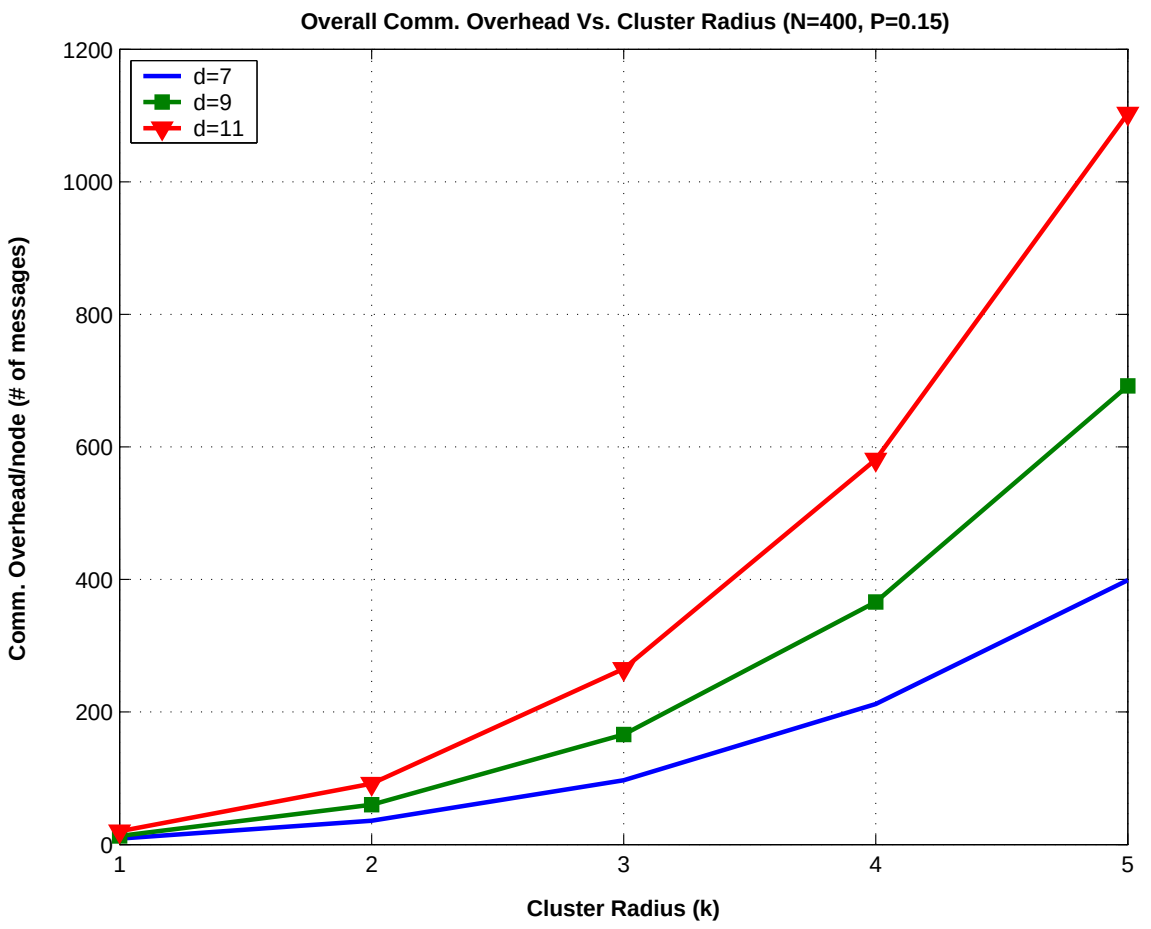

(c) Effect of cluster radius ($k$)

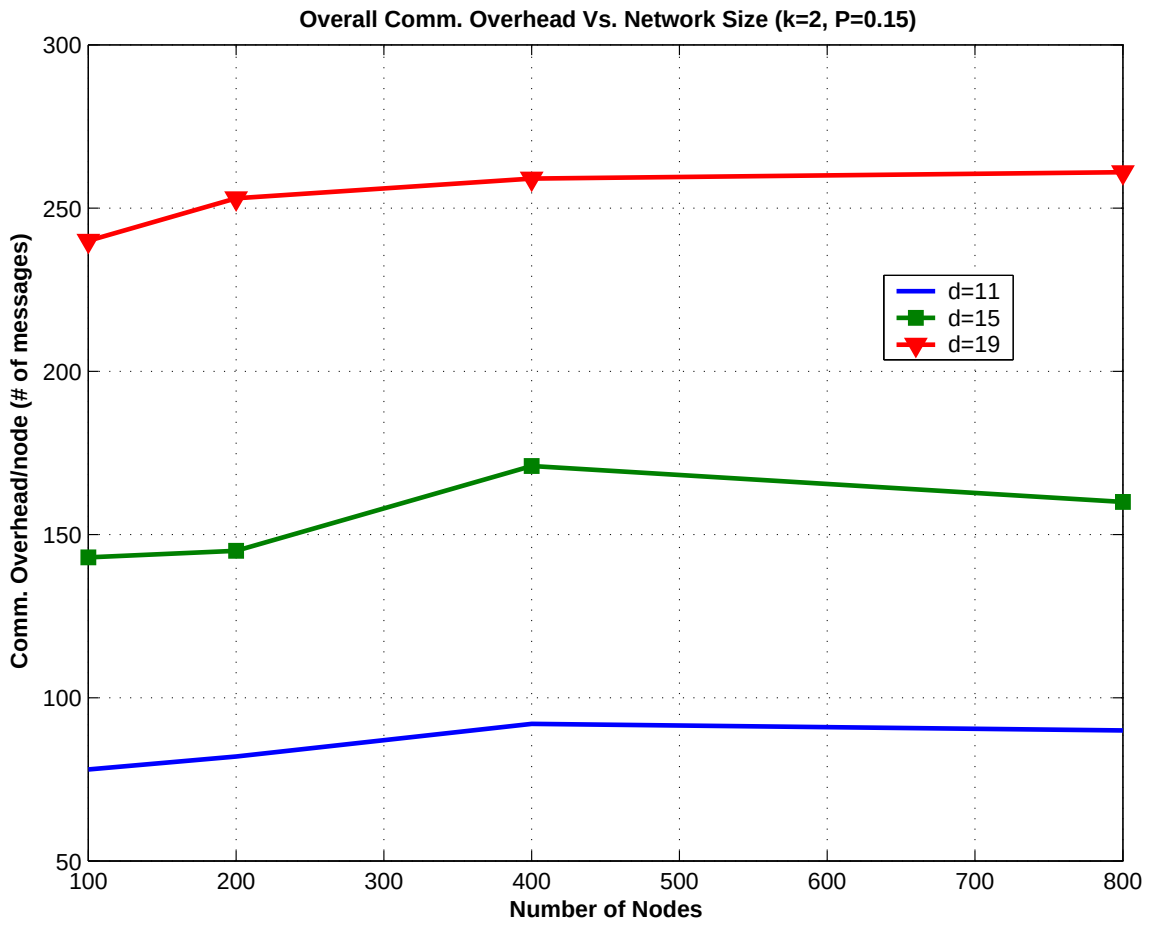

(d) Effect of network side ($n$) on comm. overhead per node

Fig. 9. The effect of different parameters on communication overhead.

Fig. 7(c) shows that the normalized stdev of the overlapping degree is always less than 4%, regardless of the node degree ($d$) or cluster radius ($k$). This means that the AOD is consistent between different clusters, and hence *KOCA* can be used to achieve a certain minimum overlapping degree as discussed in Section V-I.

### F. *Effect of the Parameters on Cluster Size*

Since clusters are overlapping, increasing the number of clusters will not affect the cluster size. Therefore, $p$ has no effect on $N_c$ as shown in Fig. 8(a). On the other hand, $N_c$ increases linearly with $d$ and quadratically with $k$ , as shown in Fig. 8(b). This confirms our analysis.

As a measure of load balancing, Fig. 8(c) shows the *normalized* stdev of the average number of nodes per cluster. The results show very low normalized stdev, less than 1.35%, regardless of the values of $d$ and $k$. This means that the *KOCA* protocol produces equal-sized clusters.

### G. *Effect of the Parameters on Communication Overhead*

Fig. 9 shows the impact of different simulation parameters on communication overhead. The numbers presented by the figures capture the communication overhead in both the cluster head advertisement phase (CHAD phase) and the join request phase (JREQ phase). We observe that the communication overhead increases linearly with $p$, cubically with ($k$), and linearly with $d$. The number of messages transmitted per node is independent from the network size. This confirms our analytical results in Section IV-C.

### H. *Contention-based MAC Protocol*

Although *KOCA* is independent from the underlying MAC protocol, its performance may be affected by the choice of a particular protocol. This section studies the performance of *KOCA* over the 802.11 MAC protocols which implements the *CSMA/CA* algorithm. In addition, we study the effect of communication errors on the protocol. For space constraints, we show only some of the results here. The reader is referred to [50] for more details.

*1) Percentage of Covered Nodes:* Figure 10 shows the effect of contention and communication error on the percentage of covered nodes. We can see from the figure the effect of these factors is limited (less than 8% for very severe communication errors of 10%). Moreover, the choice of the *KOCA* parameters ($k$, $p$, and $d$) can be adjusted to compensate for this limited drop in performance as discussed before.

*2) Average Overlapping Degree:* Figure 11 shows the effect of contention and communication error on the average overlapping degree. The figure shows that, again, the effect of these factors is limited. In addition, the choice of the *KOCA* parameters can be adjusted to compensate for this limited drop in performance.

*3) Connectivity Ratio:* Figure 13 shows the effect of contention and communication error on the connectivity ratio. For the contention-free case, the PER has no effect on the connectivity ratio. For the contention-based case, the PER effect on performance is a step function. For a range of PER, the connectivity ratio remains the same. When a certain PER is reached, the connectivity ratio drops to the second level and remains there until the next threshold. The choice of the *KOCA* parameters can be adjusted to compensate for this limited drop in performance. For the contention-free case, we expect to have the same step function performance, although over higher PER levels due to the more deterministic nature of the *TDMA* protocol.

*4) Average Cluster Size:* Figure 12 shows the effect of contention and communication error on the average cluster size. The figure shows that the average cluster size is inversely proportional with PER. For high PER, of 0.02, the average cluster size is decreased by less than 7%. Again, the choice of the *KOCA* parameters can be adjusted to compensate for this limited drop in performance.

*5) Communication Overhead:* Figure 14 shows the communication overhead (total number of messages) of *KOCA* under different packer error rates (PER) for the contention-free *TDMA* protocol and the 802.11 protocol. The figure shows that as the packet error rate increases, the communication overhead decreases. This can be attributed to the decreased traffic due to the dropped packets. The decrease in the communication overhead is linear with the increase in PER. This linear relation is because PER represents the fraction of packets that are dropped from the total number of packets. The results also show that using a contention-based algorithm reduces the communication overhead, due to the same reason of reducing the amount of traffic.

### *I. Summary*

We have shown in this section that *KOCA* satisfies the three conditions, defined in section II. The cluster head probability ($p$) plays an important role in terms of coverage and connectivity between clusters. The average node degree ($d$) and the cluster radius ($k$) can be tuned to achieve a reasonable average overlapping degree between clusters, regardless of $p$. Moreover, the stdev of the overlapping degree is less than 4% for different parameters. This means that the AOD is consistent among the various overlapping clusters.

Under contention and severe communication errors, up to 10%, *KOCA* communication overhead is reduced due to the dropped packets. Other performance parameters are either slightly affected such as the AOD and the percentage of covered nodes, or have limited performance degradation, such as the connectivity ratio and average cluster size. For the latter case, adjusting the *KOCA* parameters, mainly $k$, $d$, and $p$, can help in compensating for this limited performance degradation.

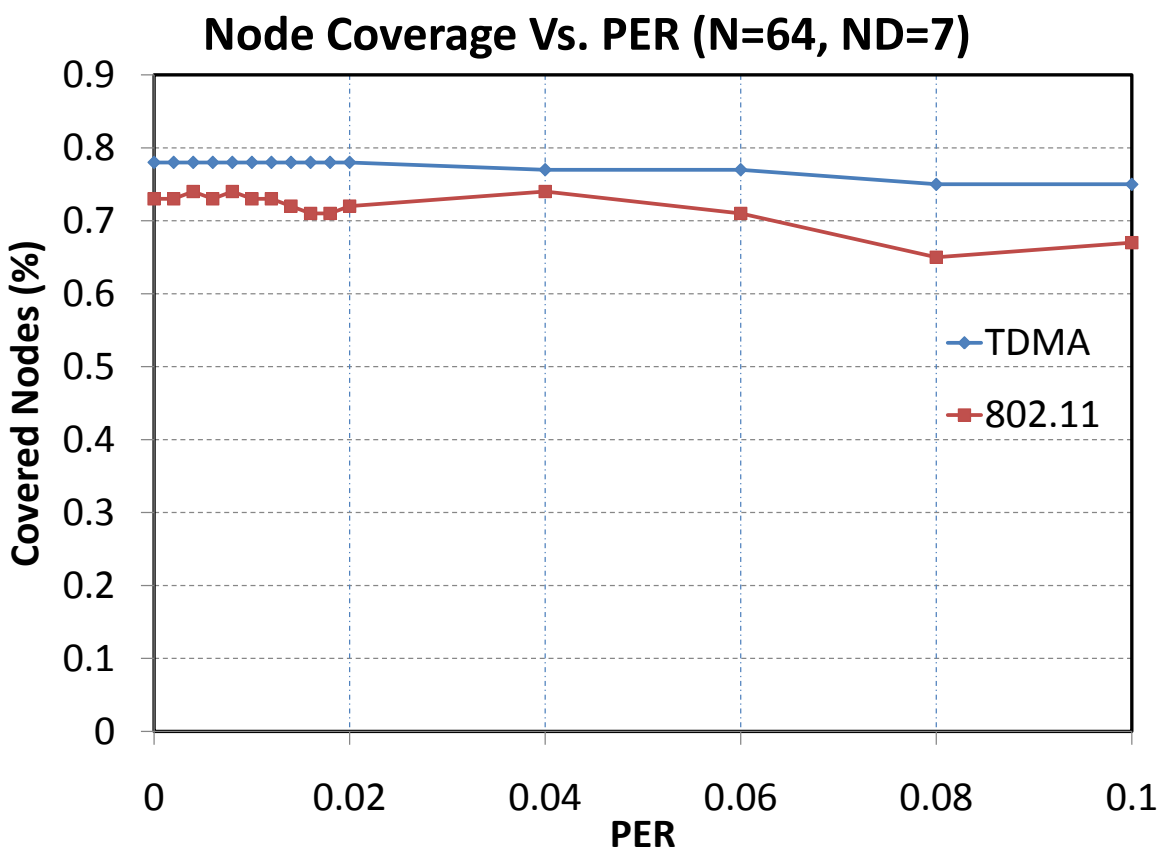

Fig. 10. Effect of contention and communication errors on percentage of covered nodes.

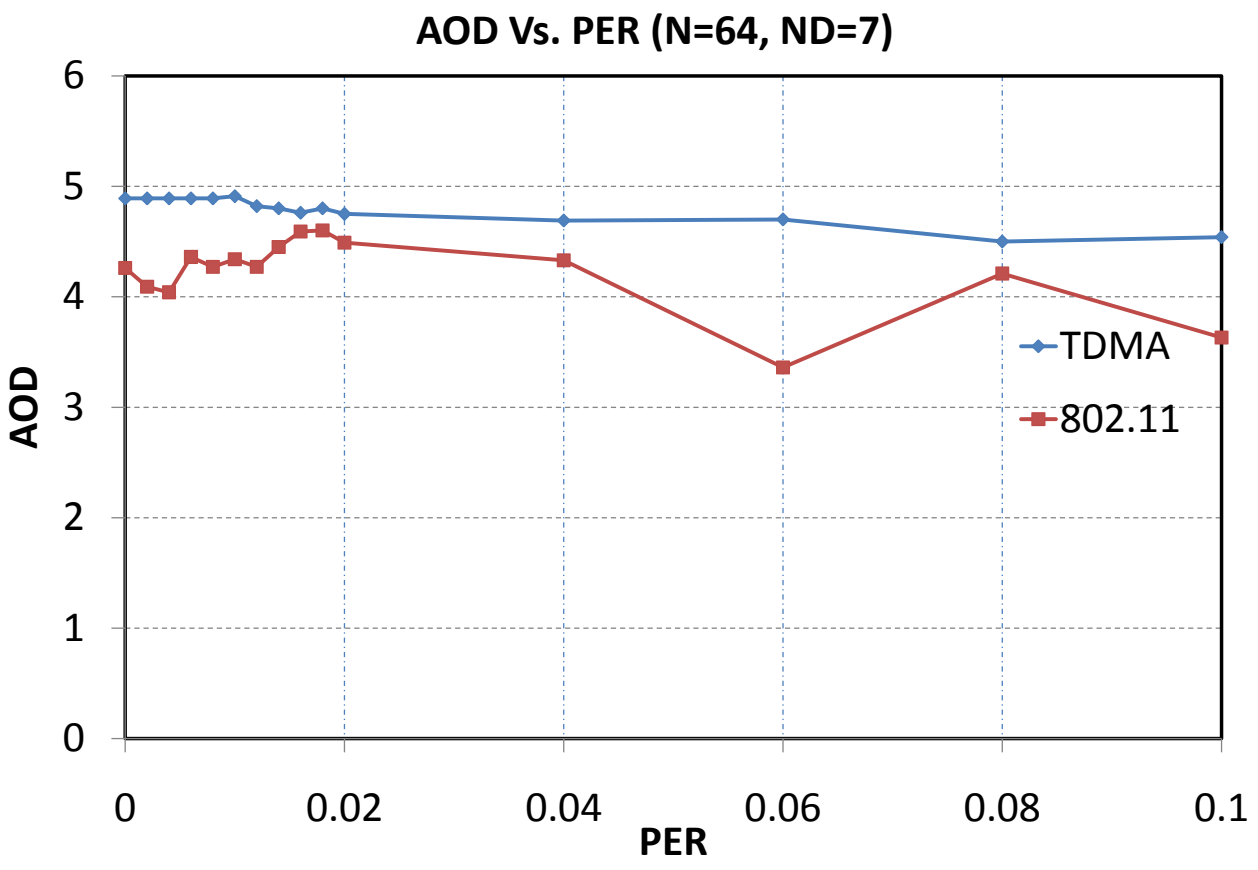

(a) Average overlapping degree

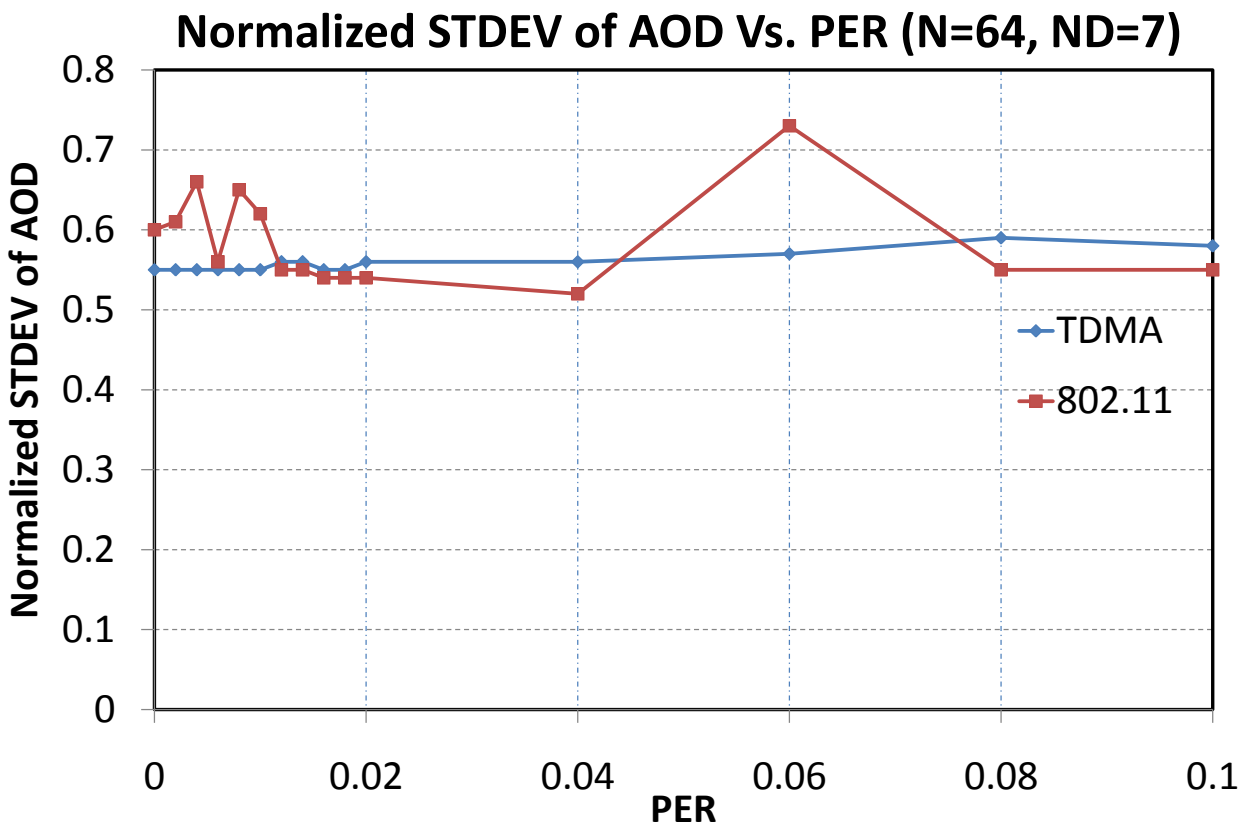

(b) Normalized STDEV of AOD

Fig. 11. Effect of contention and communication errors on overlapping degree.

Although we can select parameter values to achieve a certain average overlapping degree, and hence average number of boundary nodes between any two clusters, all the boundary nodes need not be active at the same time. This decision is based on the higher level protocols that runs on top of *KOCA*. Some applications may choose to activate only one boundary node and switch to other boundary nodes when critical events occur, e.g. boundary node failure or energy threshold reached. Another possibility to limit the number of boundary nodes, if needed, is to make each node accepts the CH_AD message with a certain probability that can be determined, similar to the analysis we presented, to reduce the number of boundary nodes.

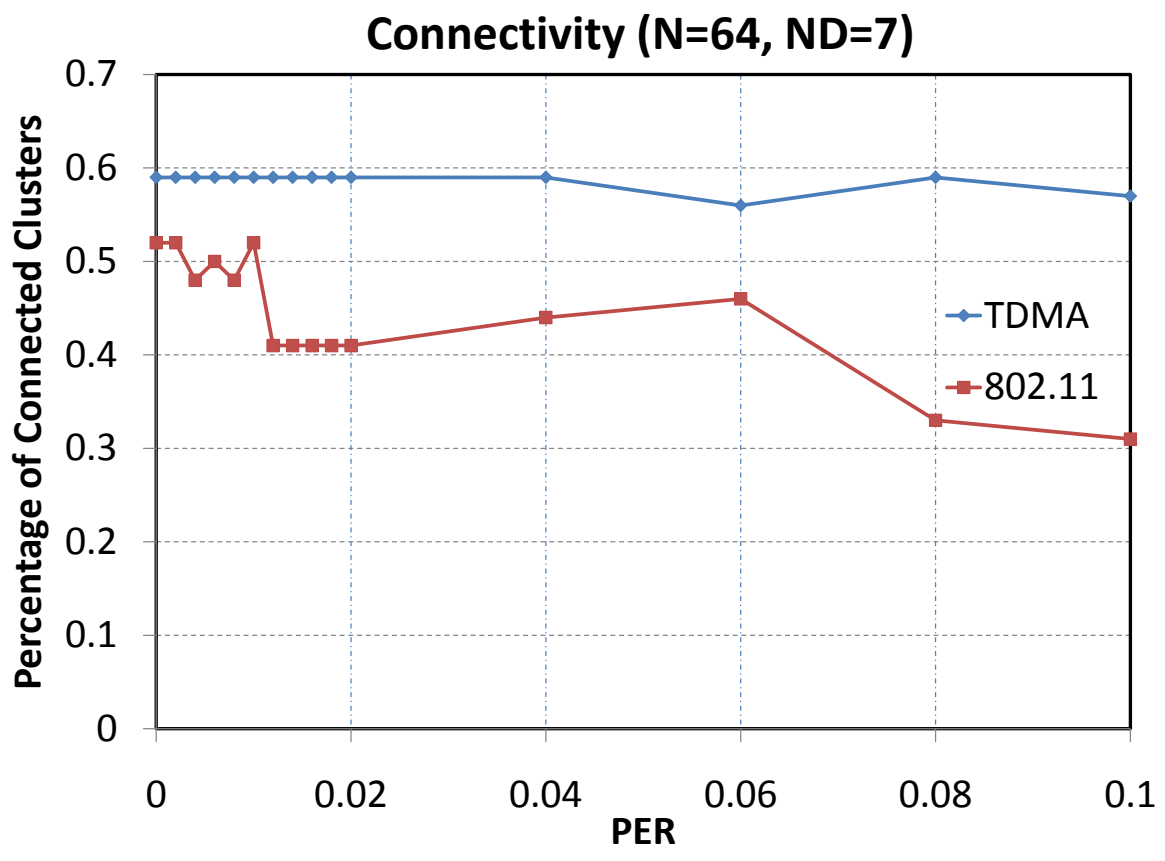

Fig. 12. Effect of contention and communication errors on connectivity ratio.

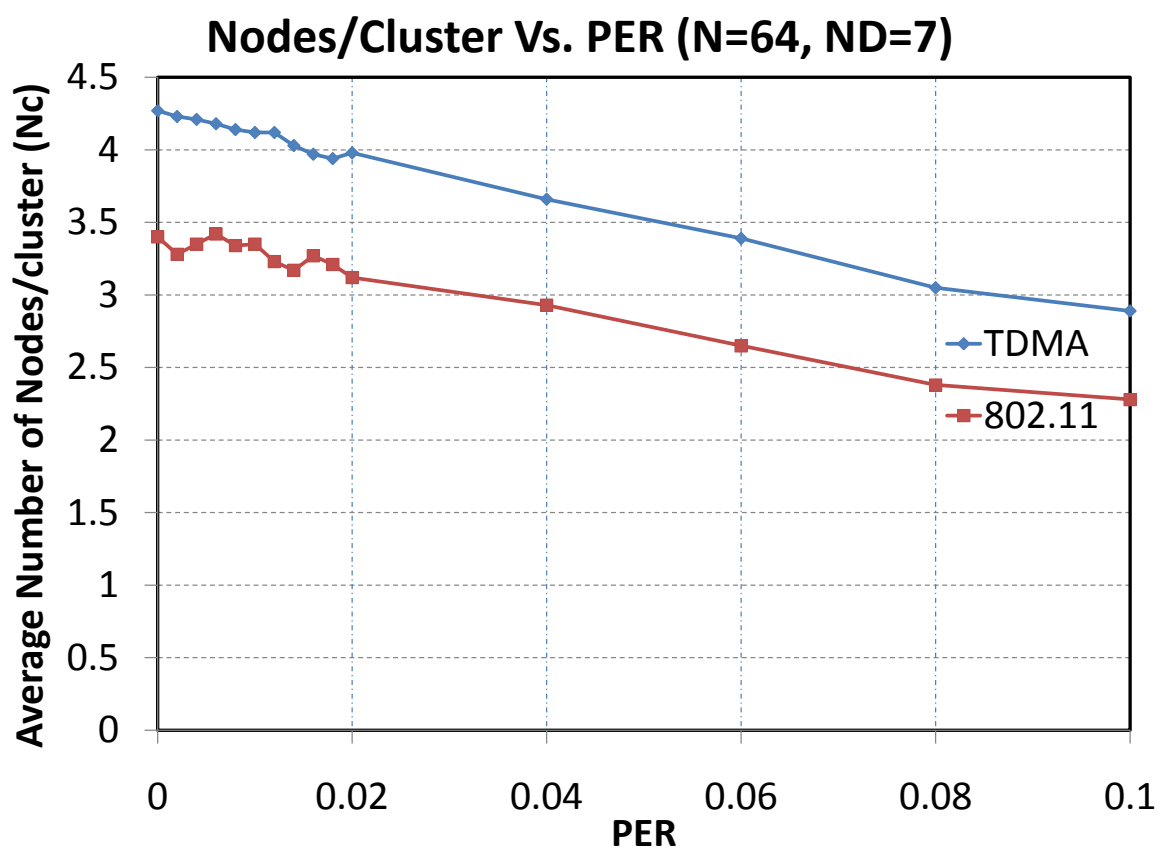

Fig. 13. Effect of contention and communication errors on average cluster size.

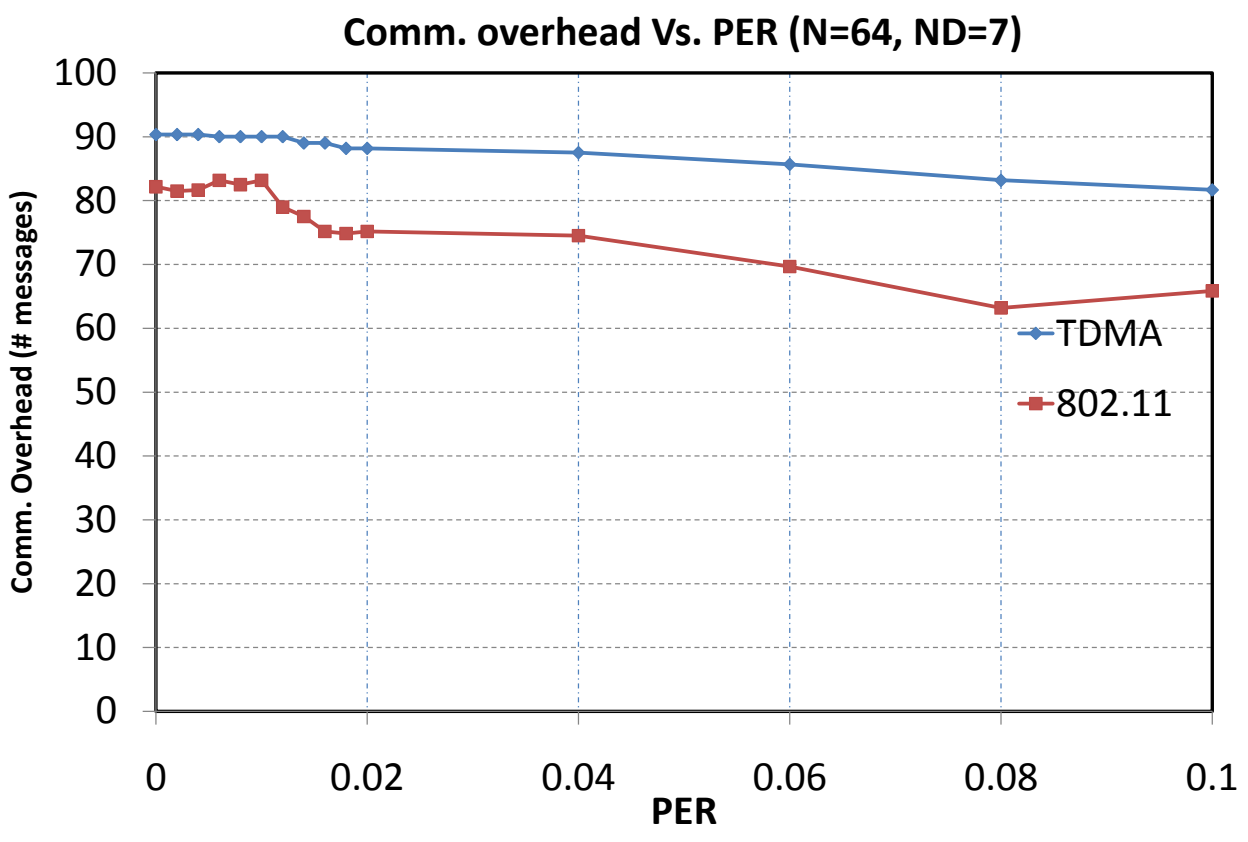

(a) Overall communication overhead

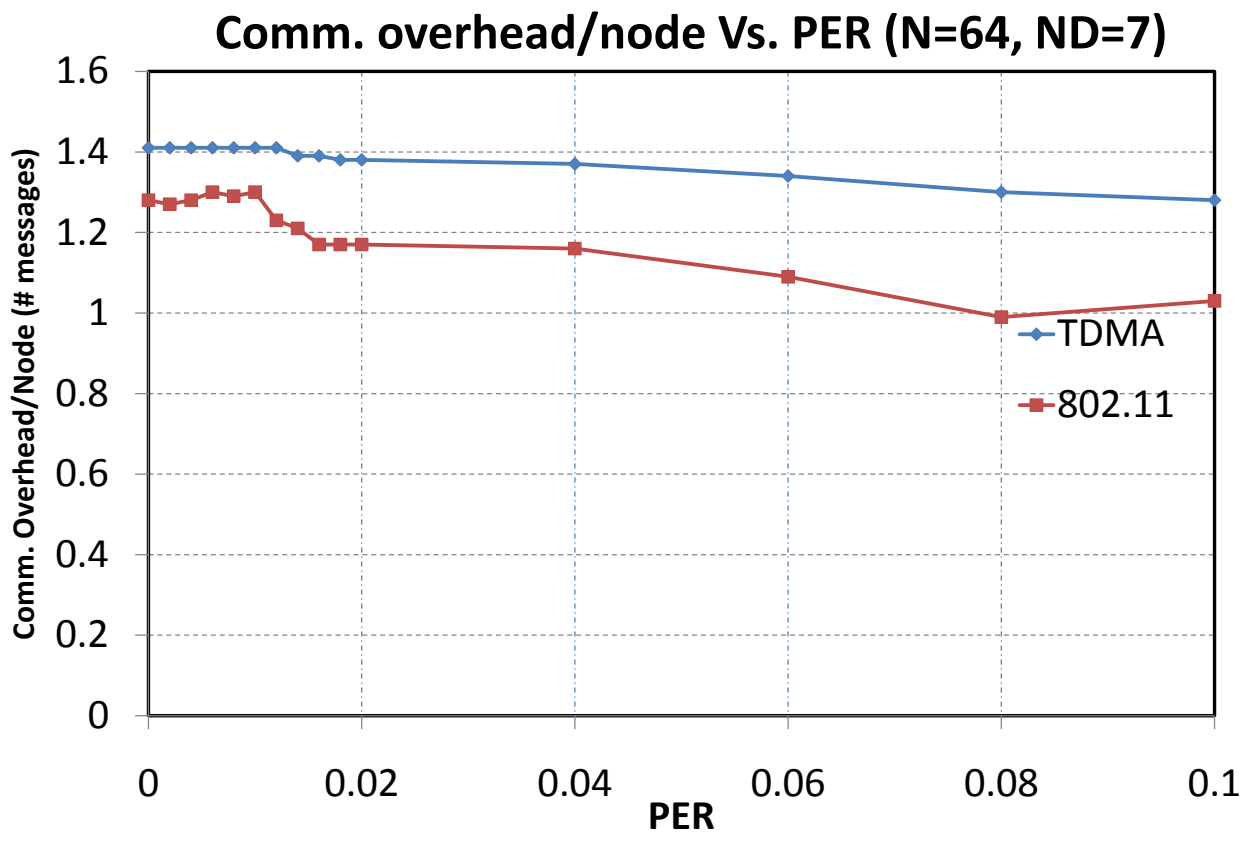

(b) Communication overhead per node

Fig. 14. Effect of contention and communication errors on communication overhead.

We have also shown that *KOCA* generates equal-sized clusters. Equal-sized clusters is a desirable property because it enables an even distribution of control (e.g., data processing, aggregation, storage load) over cluster heads; no cluster head is overburdened or under-utilized. Moreover, the results show that the average cluster size can be controlled by tuning the average node degree ($d$) or the cluster radius ($k$).

The number of messages transmitted per node due to *KOCA* is independent from the network size. We also validated our analysis in Section IV.

For some applications, limiting the cluster size is important to reduce latency, e.g. in routing protocols, or to increase accuracy, e.g. in location determination systems [29]. For applications that require high reliability, increasing the average overlapping degree is more important. On the other hand, for energy critical applications, controlling the communication overhead and transmission power may be of higher importance. The curves presented in this section can be used as guidelines for designing the sensor network for these different objectives.

The results presented show that *KOCA* can guarantee an *average* overlapping degree. Although this may be sufficient for some applications, a number of applications may require a guarantee on the *minimum* overlapping degree. One can increase the probability of achieving a certain *minimum* overlapping degree by selecting proper values for the parameters that increase the *average* overlapping degree. Our hypothesis is that increasing the average overlapping degree leads to increasing the minimum overlapping degree. This is supported by the low normalized stdev values of the overlapping degree achieved by *KOCA* (Fig. 7(c)). This still needs to be confirmed through further analysis.

## VI. Related Work

In the last few years, many algorithms have been proposed for clustering in wireless ad-hoc networks [2]–[12], [14]–[23], [25]–[27]. Clustering algorithms can be classified as either deterministic or randomized. Deterministic algorithms, e.g. [2]–[8], use weights associated with nodes to elect cluster heads. These weights can be calculated based on node degree [5], [8], node ID [2]–[4], residual energy, and mobility rate [6]. Each node broadcasts the calculated weight and a node is elected as a cluster head if it is the highest weight among its neighboring nodes. In randomized clustering algorithms, the nodes elect themselves as cluster heads with some probability $p$ and broadcast their decisions to neighbor nodes [9]–[13]. The remaining nodes join the cluster of the cluster head that requires minimum communication energy. The probability $p$ is an important parameter in a randomized algorithm. It can be a function of node residual energy [9] or hybrid of residual energy and a secondary parameter [10]. In [11], the authors obtain analytically the optimal value for $p$ that minimizes the energy spent in communication. In *KOCA*, the probability $p$ is tuned to control the number of overlapping clusters in the network.

The DCA [5] elects the node that has the highest node degree among its 1-hop neighbors as the cluster head. It is suitable for networks in which nodes are static or moving at a very low speed. The DMAC [8] modifies the DCA algorithm to allow node mobility during or after the cluster set-up phase. A similar approach is used in the DACA algorithm [14]. The WCA [6] calculates the weight based on the number of neighbors, transmission power, battery-life and mobility rate of the node. In the LCA [2], a node becomes the cluster head if it has the highest identity among all nodes within one hop of itself or among all nodes within one hop of one of its neighbors. The LCA algorithm was revised [3] to decrease the number of cluster heads produced in the original LCA. Both LCA and LCA2 heuristics were developed to be used with small networks of less than 100 nodes.

Many of these clustering algorithms, e.g. [6]–[8], are designed with an objective of generating stable clusters in environments with mobile nodes. However, in a typical wireless sensor network, sensors' locations are fixed. However, the network is still dynamic because of nodes' failure or adding new nodes. Moreover, the clustering time complexity in some protocols, e.g. [5]–[7] is O($n$), where $n$ is the total number of nodes in the network. This makes them less suitable for sensor networks that have a large number of sensors. Unlike those protocols, *KOCA* terminates in a constant number of iterations. Some clustering algorithms make assumptions about node capabilities, e.g., location-awareness [22]–[24] or clock synchronization among the nodes [2], [3]. This is again not a reasonable assumption in case of low-cost low-power sensor networks.

The majority of clustering algorithms construct clusters where every node in the network is no more than 1-hop away from a cluster head [5], [8]–[10], [26]. We call these *single-hop* clusters. For example, the HEED [10] algorithm forms single-hop clusters with the objective of prolonging network lifetime. In [9], Heinzelman et al. have proposed a distributed algorithm for wireless sensor networks (LEACH) in which sensors randomly elect themselves as cluster heads with some probability and broadcast their decisions. The remaining sensors join the cluster of the cluster head that requires minimum communication energy. Similarly, Baker et al [2] construct overlapping cluster with $k = 1$. In large networks single-hop clustering may generate a large number of cluster heads and eventually lead to the same problem as if there is no clustering.

Few papers have addressed the problem of multi-hop ($k$-hop) clustering [4], [11], [13]. These algorithms are mostly heuristic in nature and aim at generating the minimum number of disjoint clusters such that any node in any cluster is at most $k$ hops away from the cluster head. In [4], the authors presented the Max-Min heuristic to form non-overlapping $k$-clusters in a wireless ad hoc network. The Max-Min algorithm does not ensure that the energy used in communicating information to the information center is minimized. In [11], the authors proposed a LEACH-like randomized clustering algorithm for organizing the sensors, in a wireless sensor network, in a hierarchy of clusters with an objective of minimizing the energy spent in communicating the information to the processing center. Their main focus was to find the optimal number of cluster heads at each level of clustering analytically, and apply this recursively to generate one or more levels of clustering. However, our main focus is to generate overlapping clusters with certain overlapping degree.

To the best of our knowledge, this is the first paper to discuss the problem of overlapping multi-hop clustering.

## VII. Conclusion

In this paper, we have formulated the overlapping multi-hop clustering problem for wireless sensor networks that appears in many sensor network applications. Since the problem is NP-hard, we have introduced the *KOCA* randomized multi-hop heuristic algorithm that generates connected overlapping clusters covering the entire sensor network with a specific average overlapping degree.

We have studied the characteristics of *KOCA* through analysis and simulation. The results indicate that *KOCA* provides high network coverage and connectivity. Moreover, by selecting the parameter values we can achieve a certain average overlapping degree and control the cluster size. The overlapping degree has a low stdev which provides a consistent overlap between different clusters. In addition, *KOCA* terminates in a constant number of iterations independent of the network size.

Although *KOCA* generates overlapping clusters, the simulation results show that the clusters are approximately equal in size. This is desirable to achieve load balancing between different clusters.

## Appendix A
## KOCA Algorithm

Fig. 15 shows the activities of the *KOCA* clustering algorithm using an event-based notation. In this appendix, we show that *KOCA* meets the following design goals (requirements):

1) Completely distributed.
2) Terminates within O($k$) iterations, regardless of network diameter, where $k$ is the cluster radius.
3) At the end of the algorithm, each node is either a cluster head, or non-cluster head node that belongs to one or more clusters.
4) Efficient in terms of memory usage.

**Requirement 1** *KOCA* is completely distributed: A node can either elect to become a cluster head, or join a cluster if it receives *CH_AD* messages within its cluster radius. Thus, node decisions are based solely on local information.

**Lemma A.1** (Requirement 2)**.** *The time complexity of* KOCA *is O($k$).*

*Proof:* The worst case running time occurs when a non-CH (NCH) node does not receive any *CH_AD* messages and changes its status to CH. Then, this node broadcasts *CH_AD* message and waits for JREQ messages. Recall from Section III-C that the maximum time that an NCH node waits for *CH_AD* message is equal to $t(k) + \delta$, where $t(k)$ is the time needed for a message to travel $k$ hops and $\delta$ is a constant value independent from $k$. Hence, the total time of this worst case scenario is $t(k) + \delta + 2t(k)$. Therefore the maximum time that a node should wait before terminating *KOCA* is $t(k) + \delta + 2t(k) = 3t(k) + \delta = O(k)$. ■

**Lemma A.2** (Requirement 3)**.** *At the end of the* KOCA *algorithm, a node is either a cluster head, or non-cluster head node that belongs to one or more clusters.*

*Proof:* Initially each node is either CH or NCH node. If the node is a CH node, it will terminate the *KOCA* algorithm after $2t(k) + \delta$ time units when the *JREQ_WAIT* timer fires. In case of NCH node, after $t(k) + \delta$ time units, either it joins one or more of the clusters that it heard from or change status to CH and terminates the *KOCA* algorithm after $2t(k)$ time units. ■

**Lemma A.3.** *The expected number of adjacent overlapping clusters is $O(pdk^2)$, where $p$ is the cluster head probability, $d$ is the average node degree, and $k$ is the cluster radius.*

*Proof:* Recall that the expected number of clusters is $np$ where $n$ is the network size. Let $u$ and $v$ be two cluster head nodes. Then the two corresponding clusters of $u$ and $v$ are overlapping $\iff$ $dist_G(u,v) < 2k$. Using the circle approximation of the cluster as discussed in Section IV, then the probability $(P_{Adj})$ that a CH node is within distance $2R, R = kT_r$, from $m$ other CH nodes is given by the following binomial distribution:

$$P_{Adj}(m) = P_{2R}^{m}(1 - P_{2R})^{np-m-1} \begin{pmatrix} np-1 \\ m \end{pmatrix} \tag{15}$$

where $P_{2R} = \frac{\pi(2R)^2}{l^2}$q

Hence, the expected number of adjacent clusters is $(E(P_{Adj}))$:

$$E(P_{Adj}) = P_{2R}(np-1) \simeq npP_{2R} = \frac{4\pi npR^2}{l^2} \tag{16}$$

Since $R = kT_r$, substituting from equation 1 and simplifying the expression, we get the following:

$$E(P_{Adj}) = 4\pi pdk^2 = O(pdk^2) \tag{17}$$

■

**Lemma A.4** (Requirement 4)**.** *The* KOCA *algorithm has an average memory usage of $O(1)$ per node.*

*Proof:* The two major data structures used by the *KOCA* protocol are: *CH_table* and *AC_table*. All other data structures will take O(1) memory to store. Recall from Section III-A, *CH_table* is used by each node, whether CH or NCH, to store information about the known CH nodes. Hence, the average size of the *CH_table* is equal to the expected number of clusters that cover a certain node; which is equal to the expected number of adjacent clusters $(E(P_{Adj}))$. Therefore, using lemma A.3, the average size of the *CH_table* is $O(dk^2)$. Since both $d$, and $k$ are constants and independent of the network size, the average size of *CH_table* is $O(1)$. Note that the maximum size of *CH_table* can not exceed the average number of clusters $(pn)$.

Recall from Section III-A, *AC_table* is used by only CH nodes to keep track of adjacent clusters. Hence, we can calculate the average size of *AC_table* as follows:

size(*AC_table*) = $E(P_{Adj})$ x *the expected number of boundary nodes*

However, the expected number of boundary nodes is equal to the average overlapping degree (AOD). Substituting from Eq.8, we get the following:

size(*AC_table*) = $E(P_{Adj})$ x $\frac{dk^2}{4\mu}$ = $O(\frac{d^2k^4}{\mu})$

Since both $d$, and $k$ are constants and independent of the network size, the average size of *AC_table* is $O(1)$. Hence, on the average, the total memory usage per node is $O(1)$. ■

```
Initialization() // executed once
ac:
r = generate random number from 0..1;
if r < p then
  status := CH;
  broadcast (CH_AD, NID, NID, 1);
  set JREQ_WAIT timer;
else
  status := NCH;
  set CH_AD_WAIT timer;

CH_AD_Received (SID, CHID, HC)
ac: if status = NCH
  if CHID is not in the CH_table
    Add (CHID, HC, SID) to CH_table;
    if HC < k
      HC := HC + 1;
      broadcast (CH_AD, NID, CHID, HC);
    // else HC ≥ k, do not forward the message more than k hops
  // else you have already heard of this cluster, do nothing
else
  // node is a CH node
  if CHID = NID
    discard the message; // This is an echo message
  if CHID is not in the AC_table
    Add (CHID, NID) to AC_table;
    Add (CHID, HC, SID) to CH_table;
    if HC < k
      HC := HC + 1;
      broadcast (CH_AD, NID, CHID, HC);
    // else HC ≥ k, do not forward the message more than k hops
  // else you have already heard of this cluster, do nothing

JREQ_Received (RID, SID, CHID, nd, (NID, RSID, NID)_{1..nd}, nc, (CHID)_{0..nc})
ac: if status = NCH
  if RID = NID
    RID := CH_table[CHID].prev;
    broadcast (JREQ, RID, SID, CHID, nd, (NID, RSID, NID)_{1..nd}, nc, (CHID, cost)_{0..nc});
  // else do nothing to limit the flooding of JREQ message
else
  // node is a CH node
  if CHID = NID
    Add SID to the set of vertices in LCG;
    Add (NID, RSID, NID)_{1..nd} to the set of edges in LCG;
    Add (CHID, cost, SID)_{0..nc} to the AC_table;
  else
    RID := CH_table[CHID].prev;
    broadcast (JREQ, RID, SID, CHID, nd, (NID, RSID, NID)_{1..nd}, nc, (CHID, cost)_{0..nc});

EndClusterFormationPhase
ec: JREQ_WAIT timer fires. // for CH node
ac: Start the Local Location Discovery (LLD) phase using information stored in LCG and AC_table.

ChangeStatus
ec: CH_AD_WAIT timer fires. // for NCH node
ac: if CH_table empty
  status := CH;
  broadcast (CH_AD, NID, NID, 1);
  set JREQ_WAIT timer;
else
  for all CHID in CH_table
    RID := CH_table[CHID].prev;
    broadcast (JREQ, RID, NID, CHID, (NID, RSID, NID)_{1..d}, (CHID)_{0..m});
```

Fig. 15. The *KOCA* Algorithm